\documentclass[fleqn,usenatbib]{mnras}
\usepackage{newtxtext,newtxmath}
\usepackage[T1]{fontenc}
\DeclareRobustCommand{\VAN}[3]{#2}
\let\VANthebibliography\thebibliography
\def\thebibliography{\DeclareRobustCommand{\VAN}[3]{##3}\VANthebibliography}
\usepackage{graphicx}	
\usepackage{amsmath}	
\usepackage{natbib}
\usepackage{multirow}
\usepackage{threeparttable}
\usepackage{xcolor}
\title[Differential reddening in front of GCs]{Differential reddening in the direction of 56 Galactic globular clusters}

\author[M.\,V.\,Legnardi et al.] 
       {M.\,V.\,Legnardi$^{1}$\thanks{E-mail: mariavittoria.legnardi@phd.unipd.it},
       A.\,P.\,Milone$^{1,2}$,      
       G.\,Cordoni$^{1}$,
       E.\,P.\,Lagioia$^{1}$,
       E.\,Dondoglio$^{1}$,
       A.\,F.\,Marino$^{2,3}$,
       \newauthor S.\,Jang$^{4}$,
       A.\,Mohandasan$^{1}$,
       and T.\,Ziliotto$^{1}$
\\ 
$^{1}$ Dipartimento di Fisica e Astronomia ``Galileo Galilei'', Univ. di Padova, Vicolo dell'Osservatorio 3, Padova, IT-35122\\
$^{2}$ Istituto Nazionale di Astrofisica - Osservatorio Astronomico di Padova, Vicolo dell’Osservatorio 5, Padova, IT-35122 \\
$^{3}$ Istituto Nazionale di Astrofisica - Osservatorio Astrofisico di Arcetri, Largo Enrico Fermi, 5, Firenze, IT-50125 \\
$^{4}$ Center for Galaxy Evolution Research and Department of Astronomy, Yonsei University, Seoul 03722, Korea
}
\date{Accepted 2023 April 05. Received 2023 April 04; in original form 2022 December 23}
\pubyear{2023}

\begin{document}
\label{firstpage}
\pagerange{\pageref{firstpage}--\pageref{lastpage}}

\maketitle

\begin{abstract}
The presence of differential reddening in the direction of Galactic globular clusters (GCs) has proven to be a serious limitation in the traditional colour-magnitude diagram (CMD) analysis. Here, we estimate local reddening variations in the direction of 56 Galactic GCs. To do that, we use the public catalogs derived as part of the {\it Hubble Space Telescope} UV Legacy Survey of Galactic Globular Clusters, which include photometry in the F275W, F336W, F438W, F606W, and F814W filters. We correct photometry for differential reddening finding that for 21 out of 56 GCs the adopted correction procedure significantly improves the CMDs. Moreover, we measure the reddening law in the direction of these clusters finding that $R_{V}$ exhibits a high level of variability within the Galaxy, ranging from $\sim2.0$ to $\sim4.0$. The updated values of $R_{V}$ have been used to improve the determination of local reddening variations and derive high-resolution reddening maps in the direction of the 21 highly-reddened targets within our sample. To compare the results of the different clusters, we compute the 68$^{\rm th}$ percentile of the differential-reddening distribution, $\sigma_{\Delta A_{\rm F814W}}$. This quantity ranges from 0.003 mag to 0.030 mag and exhibits a significant anti-correlation with the absolute module of the Galactic latitude and a strong correlation with the average reddening in the direction of each cluster. Therefore, highly-reddened GCs located in the proximity of the Galactic plane typically show higher differential-reddening variations across their field of view.
\end{abstract}


\section{Introduction}
\label{sec:intro}
Galactic globular clusters (GCs) are remarkable fossils from the early Universe. Being among the most ancient objects of the Galaxy, they provide unique insights into nucleosynthesis, star formation, and chemical enrichment at the earliest stages of the Galaxy's formation. One of the most serious challenges in the study of many GCs, especially the ones in the direction of the Galactic Center, is the presence of spatially-variable extinction, better known as differential reddening, caused by variation of dust column density across the field of view (FoV). By introducing a significant broadening of all evolutionary sequences in colour-magnitude diagrams (CMDs), differential reddening complicates the traditional CMD analysis preventing an accurate determination of some fundamental parameters, such as age, distance, and metallicity \citep[e.g.][]{bonatto2013}. Moreover, differential reddening hinders the detection and characterization of the various stellar sequences that compose the CMDs, including binaries and multiple stellar populations in GCs \citep[e.g.][]{milone2012a}. Therefore, correcting the photometry for the effects of differential reddening is a crucial step for a correct analysis of the CMD.
   
Over the years, several methods have been developed to model and correct for spatially variable extinction in the direction of Galactic GCs. A first approach consists of dividing the FoV across the cluster in a regular cell grid, extracting the CMD of the stars included in each cell, and computing the differential reddening of each cell as the average colour distance, calculated across the reddening line, between the extracted CMD and the bluest ones \citep[e.g.][]{kaluzny1993, heitsch1999, piotto1999, vonbraun2001, sarajedini2007, mcwilliam2010, gonzalez2011, bonatto2013}. An alternative method has been introduced by \cite{milone2012a} and then extensively used to investigate star clusters in the Milky Way and in Magellanic Clouds \citep[e.g.][]{lagioia2014a, milone2015, bellini2017a, li2017a, pallanca2021, dondoglio2022, legnardi2022}. In a nutshell, they estimated the differential reddening of each star in a cluster by defining a reference line for the upper main sequence (MS), the sub-giant branch (SGB), and the red-giant branch (RGB) and then calculating the colour distance of each cluster star from it along the reddening vector. The final differential-reddening value associated with each star corresponds to the median colour displacement of the spatially closest objects. 

Recently, our group has undertaken an extensive investigation of differential reddening in the FoV of more than a hundred star clusters in the Magellanic Clouds by means of {\it Hubble Space Telescope} ({\it HST}) images \citep{milone2022a}. Moreover, we investigated reddening variations across the FoV of 43 Galactic GCs \citep{jang2022}, based on ground-based wide-field photometry from \cite{stetson2019a}. Here, we constrain the amount of differential reddening in the central regions of 56 Galactic GCs by using high-precision {\it HST} photometry and astrometry by \citet{nardiello2018}. 

\cite{milone2012a} and \cite{bonatto2013}  investigated  local reddening variations across the FoV for a large sample of Galactic GCs (59 and 66, respectively) by using {\it HST} data. Both works are based on the data of the GO-10775 program alone (PI: A.\,Sarajedini; \citealt{sarajedini2007}), which is a {\it HST} Treasury project where 66 GCs have been homogeneously observed through the F606W and F814W filters of the Wide Field Camera of the Advanced Camera for Surveys (WFC/ACS). Recently, this dataset has been integrated through the images taken as part of the GO-13227 program (PI: G.\,Piotto; \citealt{piotto2015}), a UV-initiative proposal designed to observe a large sample of clusters through the UV/blue F275W, F336W, and F438W bands of the Ultraviolet and Visual Channel of the Wide Field Camera 3 (UVIS/WFC3) on board {\it HST}. 

Based on multi-band photometry, we estimate the differential reddening associated with each source in the catalog and derive a high-resolution and high-precision reddening map in the directions of 56 Galactic GCs. To do that, we adapt to the available catalogs \citep{nardiello2018} the approach by \cite{milone2012a}. The main difference is that our dataset comprises magnitudes in five filters, namely F275W, F336W, F438W, F606W, and F814W, whereas the work by Milone and collaborators is based only on the F606W and F814W bands. Our procedure, combining information from five magnitudes, provides more accurate differential-reddening corrections than the previous one.  Moreover, we take advantage of stellar proper motions that allow us to separate the bulk of cluster members from field stars in all GCs. The resulting differential-reddening catalogs and the reddening maps are publicly released to the astronomical community.  

This paper is organized as follows. Section\,\ref{sec:data} presents the GC sample and describes the  procedures used to identify those stars that provide information on differential reddening. In Section\,\ref{sec:dr} we explain in detail the method to calculate differential reddening, and we estimate the impact of changing the reddening law on the differential-reddening determination. In Section\,\ref{subsec:rl} we describe the procedure to constrain the reddening law in the direction of the highly-reddened clusters and we construct high-resolution reddening maps. Finally, Section\,\ref{sec:conc} briefly overviews the results.

\begin{figure} 
  \centering
  \includegraphics[width=.45\textwidth,trim={0cm 0cm 0cm 0cm},clip]{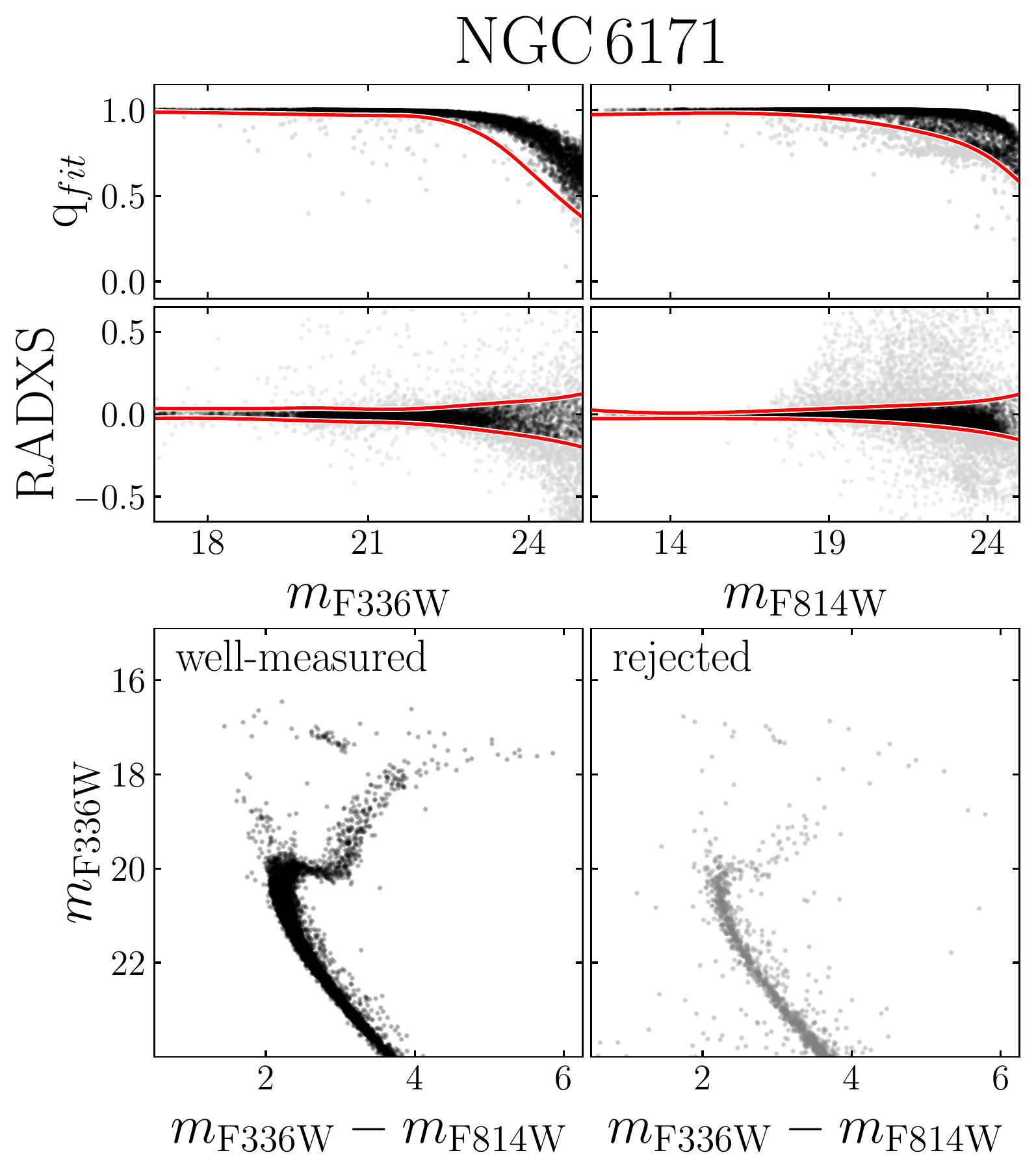}
  \caption{This figure illustrates the procedure to select well-measured stars in NGC\,6171. The top and middle panels present the selection based on the q$_{fit}$ (top) and RADXS (middle) parameters. Red lines separate well-measured stars (black points) from those that have poorer photometry (gray points). The $m_{\rm F336W}$\,versus\,$m_{\rm F336W}-m_{\rm F814W}$ CMDs for well-measured stars and for the stars that we rejected are displayed in the bottom-left and the bottom-right panel, respectively.}
  \label{fig:cmdsel}
\end{figure} 

\section{Data and data analysis}
\label{sec:data}
To estimate the amount of differential reddening we used the catalogs by \cite{nardiello2018} that provide astrometric positions, high-precision multi-band photometry, and cluster membership from proper motions in the central $\sim 2.7^{\prime} \times 2.7^{\prime}$ FoV for a large sample of 56 Galactic GCs. All the clusters were observed with WFC/ACS camera in F606W and F814W filters as part of the GO-10775 program \citep[PI: A.\,Sarajedini;][]{sarajedini2007} and with UVIS/WFC3 in F275W, F336W, and F438W filters mostly within GO-13227 \citep[PI: G.\,Piotto;][]{piotto2015}. 

The data reduction has been carried out with the computer program KS2 developed by Jay Anderson as an evolution of \texttt{kitchen\_sync}, written initially to reduce the images from GO-10775 \citep{anderson2008}. KS2 derives three distinct astro-photometric catalogs for each cluster obtained through three different methods. Method one provides accurate measurements of bright, unsaturated stars, whereas the photometry of faint stars is best measured with methods two and three \citep[see][for details]{sabbi2016a, bellini2017a, nardiello2018, milone2022a}.

In the following, we use photometry from method one. Indeed, the regions of the CMD where the colour broadening is most sensitive to differential reddening correspond to the upper MS, SGB, and the lower part of the RGB. 
        
\begin{figure*} 
  \centering
  \includegraphics[width=.9\textwidth,trim={0cm 0cm 0cm 0cm},clip]{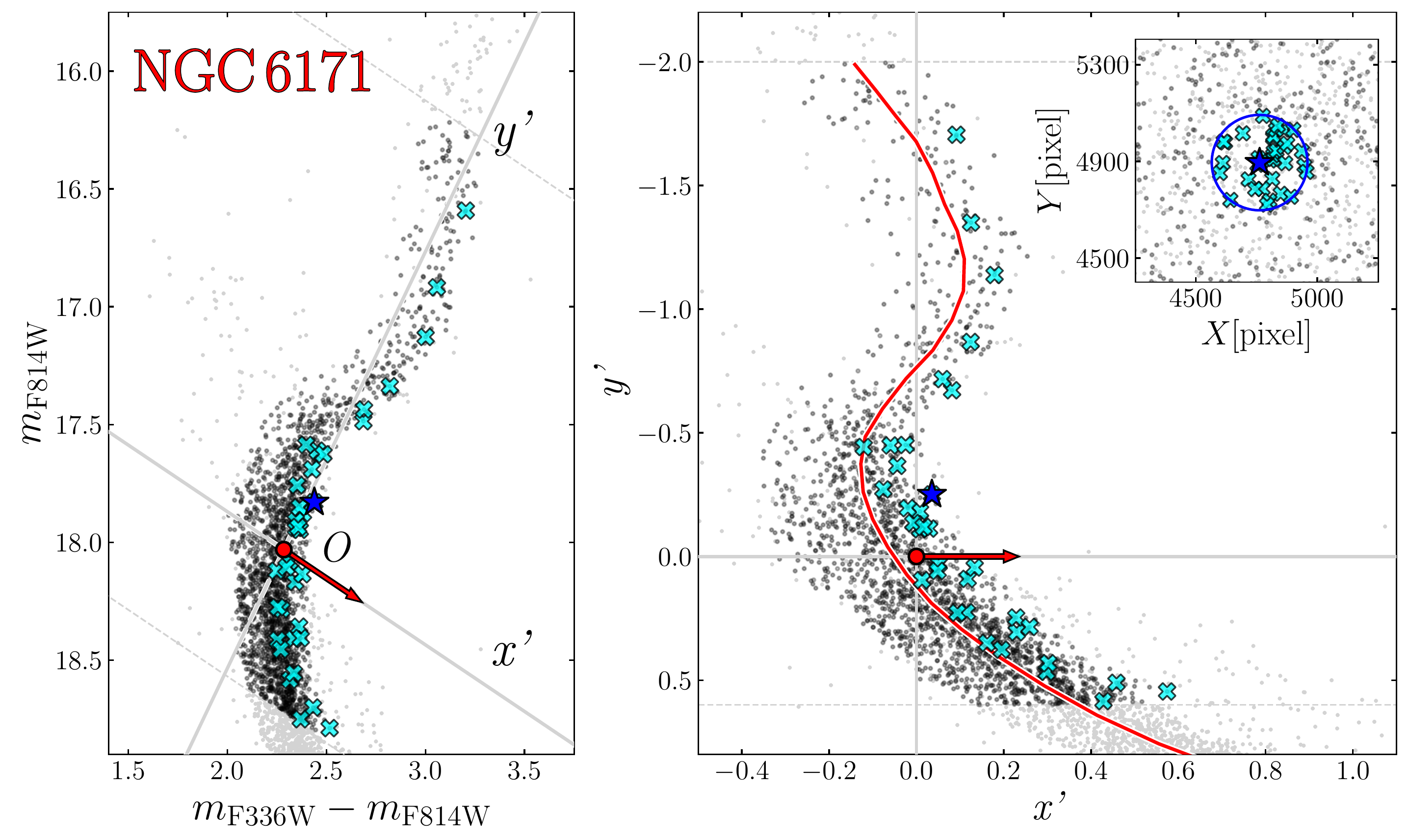}
  \caption{This figure presents the method used to calculate the differential reddening of the target star marked with the large blue starred symbol. In the left panel, we plot the $m_{\rm F814W}$ versus\,$m_{\rm F336W}-m_{\rm F814W}$ CMD, whereas in the right panel, we show the same stars but in the rotated reference frame. In both diagrams, the red arrow indicates the reddening direction, and black dots, included between the two dotted lines, identify reference stars, among which the 35 closest neighbours of the target star are represented with cyan crosses. The red line is the fiducial of reference stars calculated in the rotated reference frame. Finally, the right-panel inset shows the position of the 35 neighbouring reference stars zoomed around the target star. See Section\,\ref{sec:dr} for details on the employed procedure.} 
  \label{fig:dr_corr}
  \end{figure*} 

\subsection{Selection of well-measured cluster stars}
\label{subsec:sele}
The main impact of differential reddening on the CMD is a colour broadening of the evolutionary sequences. Specifically, the stars that are affected by a larger amount of reddening than the average value exhibit red colours and faint magnitudes relative to the fiducial line derived from all cluster members in the CMD. Conversely, stars with a small amount of reddening are shifted to the bright-blue side of the fiducial.

To estimate the amount of differential reddening associated with each star in the catalog, we need to calculate the shift along the reddening direction relative to the colour and magnitude of a sample of   nearby cluster members. However, not all the stars provide information on differential reddening. In the following, we describe the method to identify the probable cluster members with high-quality photometry.

\subsubsection{Selection of probable cluster members}
The FoVs of GCs are contaminated by a variable number of background/foreground sources. To separate cluster members and field stars, we used the cluster membership probabilities provided by \citet[][see their Section\,4 for details]{nardiello2018}, which corresponds to the probability that a star is part of the cluster based on proper motions. Specifically, we considered probable cluster members all stars with a membership probability greater than 90\%. 

\subsubsection{Selection of stars with high-quality photometry}
Similarly to differential reddening, photometric uncertainties spread the stars around the fiducial line of the CMD. Hence, it would be challenging to disentangle the effect of reddening and errors from stars with large photometric uncertainties. For this reason, selecting a sample of well-measured stars is mandatory to derive accurate reddening estimates.

For all clusters, our selection criteria rely on the quality fit (q$_{fit}$) and the RADXS diagnostic parameters, which are provided in the \cite{nardiello2018} catalogs. Specifically, the q$_{fit}$ estimates the goodness of the PSF-fitting procedure, whereas the RADXS measures the amount of flux that exceeds the predictions from the best-fitting PSF. It is a powerful tool to distinguish cosmic rays and PSF artifacts, which have negative RADXS values, and to identify background galaxies, which have $\rm RADXS>0$.

Fig.\,\ref{fig:cmdsel} shows an example of the selection procedure for NGC\,6171. In the top and middle panels, we plot the q$_{fit}$ (top) and the RADXS (middle) parameters as a function of magnitudes in the F336W (left) and F814W (right) bands. Red lines are drawn by hand with the criteria for separating well-measured stars (black points) from those that have poorer photometry (gray points). The results of the selection are illustrated in the bottom-left and -right panel, where we show the $m_{\rm F336W}$ versus\,$m_{\rm F336W}-m_{\rm F814W}$ CMD for well-measured and rejected stars, respectively.  

\section{Differential reddening determination}
\label{sec:dr}
We dealt with differential reddening by adapting to our dataset the method introduced by \cite{milone2012a}
to correct the photometry of 59 Galactic GCs for differential reddening. Our dataset, including magnitudes in the F275W, F336W, F438W, F606W, and F814W bands, allowed us to obtain improved differential-reddening determinations compared to the original ones by \cite{milone2012a}, based only on the F606W and F814W magnitudes.
 
To this aim, we exploited the $m_{\rm F814W}$ versus\,$m_{\rm X}-m_{\rm F814W}$ CMDs, where X=F275W, F336W, F438W, and F606W, and derived four distinct estimates of the amount of differential reddening suffered by each star. We then averaged together information from the different colour combinations to obtain a more accurate differential-reddening determination. In the following we summarize the adopted procedure, which is illustrated in Fig.\,\ref{fig:dr_corr} for the $m_{\rm F814W}$ versus\,$m_{\rm F336W}-m_{\rm F814W}$ CMD of NGC\,6171 that we used as a test case.
\begin{itemize}
    \item The reddening vector's direction, which is indicated by the red arrow in the left panel of Fig.\,\ref{fig:dr_corr}, has a non-zero slope with respect to CMD axes. To simplify the correction process, we started by defining a new reference frame in which the abscissa ({\it x'}) and the ordinate ({\it y'}) are parallel and orthogonal to the reddening direction, respectively. To this aim, we arbitrarily identified a rotation point near the MS turn-off (red dot) and rotated the CMD counterclockwise by an angle $\theta$, defined as:
    \begin{equation}
        \theta=\arctan \frac{A_{\rm F814W}}{A_{\rm X}-A_{\rm F814W}} 
    \end{equation}
    where A$_{\rm X}$ and A$_{\rm F814W}$ are the absorption coefficients in the X and F814W filters, respectively
    \citep{dotter2008a}. A zoom around the SGB of the resulting rotated CMD is presented in the right panel of Fig.\,\ref{fig:dr_corr}, where for completeness we also show the rotated direction of the reddening vector (red arrow) and the rotation point. The reddening vector is now parallel to the {\it x'} axis.    
    \item We then selected a sample of reference stars in the CMD region that   includes the upper MS, the SGB, and the lower portion of the RGB. The reference stars distribute along sequences where there is a wide angular variation between the direction of the fiducial line and that of the reddening vector. Hence, they allow us to disentangle the contribution of photometric errors and differential reddening to the colour broadening. As an example, the selected reference stars of NGC\,6171 are marked as black points in the $m_{\rm F814W}$ versus\,$m_{\rm F336W}-m_{\rm F814W}$ CMD and in the corresponding rotated diagram of Fig.\,\ref{fig:dr_corr}.
    \item We calculated the red fiducial line plotted in the right panel of Fig.\,\ref{fig:dr_corr}. To do that, we divided the sample of reference stars into {\it y'} bins and we calculated the median values of {\it x'} and {\it y'} for each of them. We derived the fiducial line of the selected reference stars by linearly interpolating these median points.
    \item As a final step, we computed the distance from the fiducial line along the reddening direction ($\Delta\,x'$) by subtracting from the pseudo-colour ({\it x'}) of each reference star that of the fiducial line at the same {\it y'} level.
\end{itemize}
In summary, this procedure allows us to measure the value of $\Delta\,x'$ for each reference star. These quantities will be used in Section\,\ref{subsec:DR_correction}, and  \ref{subsec:DR_lowmap} to correct the photometry for the effects of differential reddening and to investigate the reddening in front of the GCs, respectively.

\begin{figure*} 
  \centering
  \includegraphics[height=8cm,trim={0cm 0cm 0cm 0cm},clip]{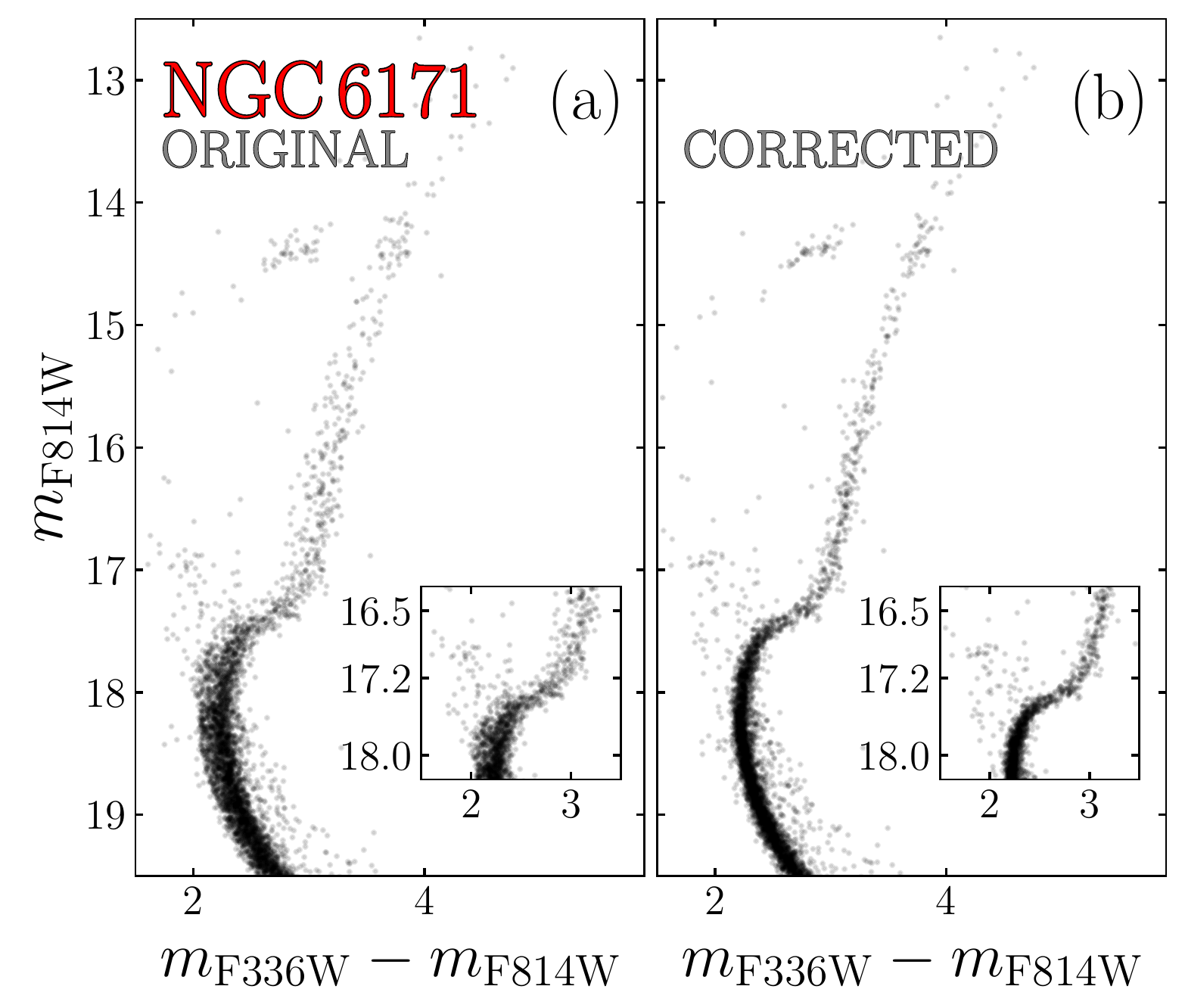}
  \includegraphics[height=8cm,trim={0cm 0cm 0cm 0cm},clip]{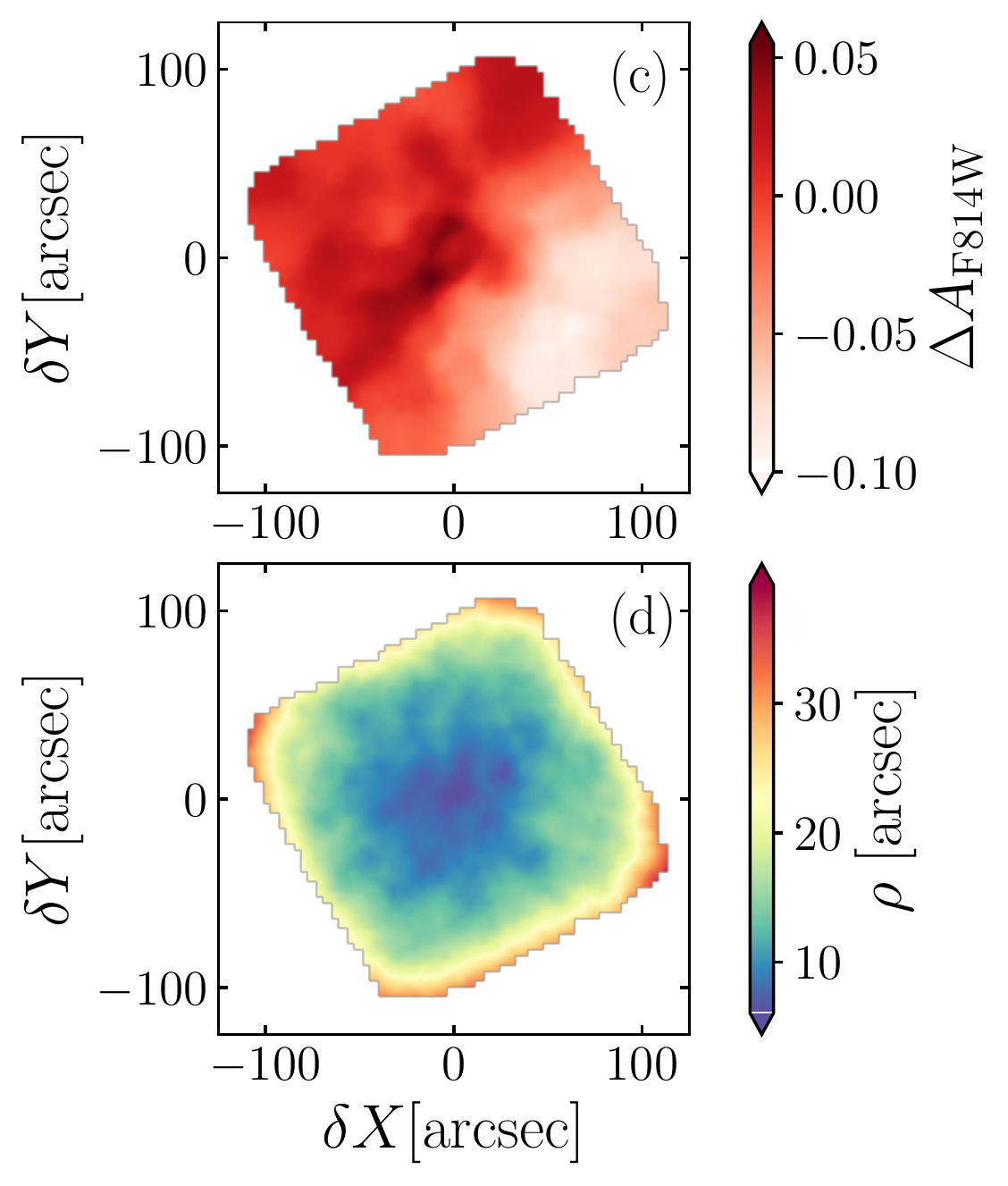}
  \caption{\textit{Panel a-b.} Comparison between the original $m_{\rm F814W}$ versus\,$m_{\rm F336W}-m_{\rm F814W}$ CMD of NGC\,6171 and the corresponding photometric diagram corrected for differential reddening. In the inset, we plotted a zoom around the SGB where the correction is more evident. \textit{Panel c.} Differential-reddening map in the direction of NGC\,6171. The levels of red are proportional to the reddening variation as indicated by the associated colour bar on the right. \textit{Panel d.} Variation of the resolution of the differential-reddening correction within the observed FoV.} 
  \label{fig:dr_ngc6171}
  \end{figure*} 

\subsection{Correcting the photometry for differential-reddening effects}
\label{subsec:DR_correction}
To derive the amount of differential reddening associated with each star in the catalog, we used an iterative procedure that consists of the following steps:
\begin{enumerate}
    \item  For each star in the catalog, we defined a local sample of $N_{\rm ref}=30$-100 nearby reference stars. 
    As an example, in Fig.\,\ref{fig:dr_corr} we mark with cyan crosses the 35 closest reference stars used to estimate the differential reddening suffered by the star represented by the blue starred symbol. As illustrated in the inset of the right panel, all the reference stars lie in a circular region centered on the target star. The radius of this circle, $\rho$, corresponds to the distance of the farthest reference star with respect to the target one, which is indicative of the map resolution in that specific position. Clearly, the value of $\rho$ changes from one star to another. It is typically small in the central cluster regions, where the density of reference stars is high, and will increase towards the external regions.  
    \item For each star, we calculated the median $\Delta\,x'$ value of the neighbouring reference-star sample excluding the target from the computation. We estimated the associated error as the root mean scatter of $\Delta\,x'$ divided by the square root of $N-1$, where $N$ is the number of selected nearby reference stars. To correct the photometry for the effects of differential reddening we subtracted the derived quantity from the $x'$ value of each star in the rotated reference frame. We then used the new diagram to select reference stars with more accuracy and therefore obtain an improved fiducial line.
    \item We repeated all steps above until the procedure converges. Typically, this happens after three iterations. At this point, we converted the corrected $x'$ and $y'$ into $m_{\rm X}$ and $m_{\rm F814W}$ magnitudes and we used the corresponding absorption coefficients to derive the differential-reddening value associated with each star.
\end{enumerate}
The best differential-reddening estimate for each star is obtained by combining the information from the different $m_{\rm F814W}$ versus\,$m_{\rm X}-m_{\rm F814W}$ CMDs. Specifically, our final estimate of $\Delta A_{\rm F814W}$ corresponds to the weighted average of the four distinct differential-reddening determinations. The corresponding uncertainties are given by the error of the weighted mean. 

We derived various determinations of differential reddening by assuming different values of $N_{\rm ref}$ ranging from 30 to 100 in step of 5. For each differential-reddening determination, we calculated the residuals between the corrected $x'$ values and the fiducial line in each CMD. The best determination of differential reddening is provided by the value of $N_{\rm ref}$  that gives the minimum values of the r.m.s of these residuals.

As an example, the results of this procedure are illustrated in panels a and b of Fig.\,\ref{fig:dr_ngc6171} for NGC\,6171, where we compare the original $m_{\rm F814W}$ versus\,$m_{\rm F336W}-m_{\rm F814W}$ CMD and the corresponding diagram corrected for differential reddening. Noticeably, the evolutionary sequences in the corrected CMD are consistently narrower than in the original one.  In panel c of Fig.\,\ref{fig:dr_ngc6171} we provide the resulting high-resolution differential-reddening map, whereas in panel d we show the variation of the resolution in the observed FoV of the cluster. As expected, the highest resolution ($\rho=5.5$ arcsec) is achieved in the central regions of the cluster where the density of the reference stars is higher. 

For 21 out of 56 GCs, the photometry corrected for differential reddening provides improved CMDs, thus demonstrating the efficiency of the correction. Conversely, the correction for differential reddening does not improve the photometry of the remaining 35 GCs. This fact is not surprising. Indeed these clusters have $E (B-V) \lesssim 0.01$ mag and the amount of differential reddening is likely comparable to or smaller than the uncertainties of $E(B-V)$ determinations.

In Table\,\ref{ebv_table} we provide for each cluster the minimum, maximum, and median value of $\rho$ used to derive the differential reddening of cluster members ($\rho_{min}$, $\rho_{max}$, and $\rho_{med}$). The average values of $\rho_{min}$, $\rho_{max}$, and $\rho_{med}$ for all clusters are 5.6, 41.9, and 11.6 arcsec, respectively.
    
\begin{figure}
    \centering
    \includegraphics[width=.4\textwidth,trim={0cm 0cm 0cm 0cm},clip]{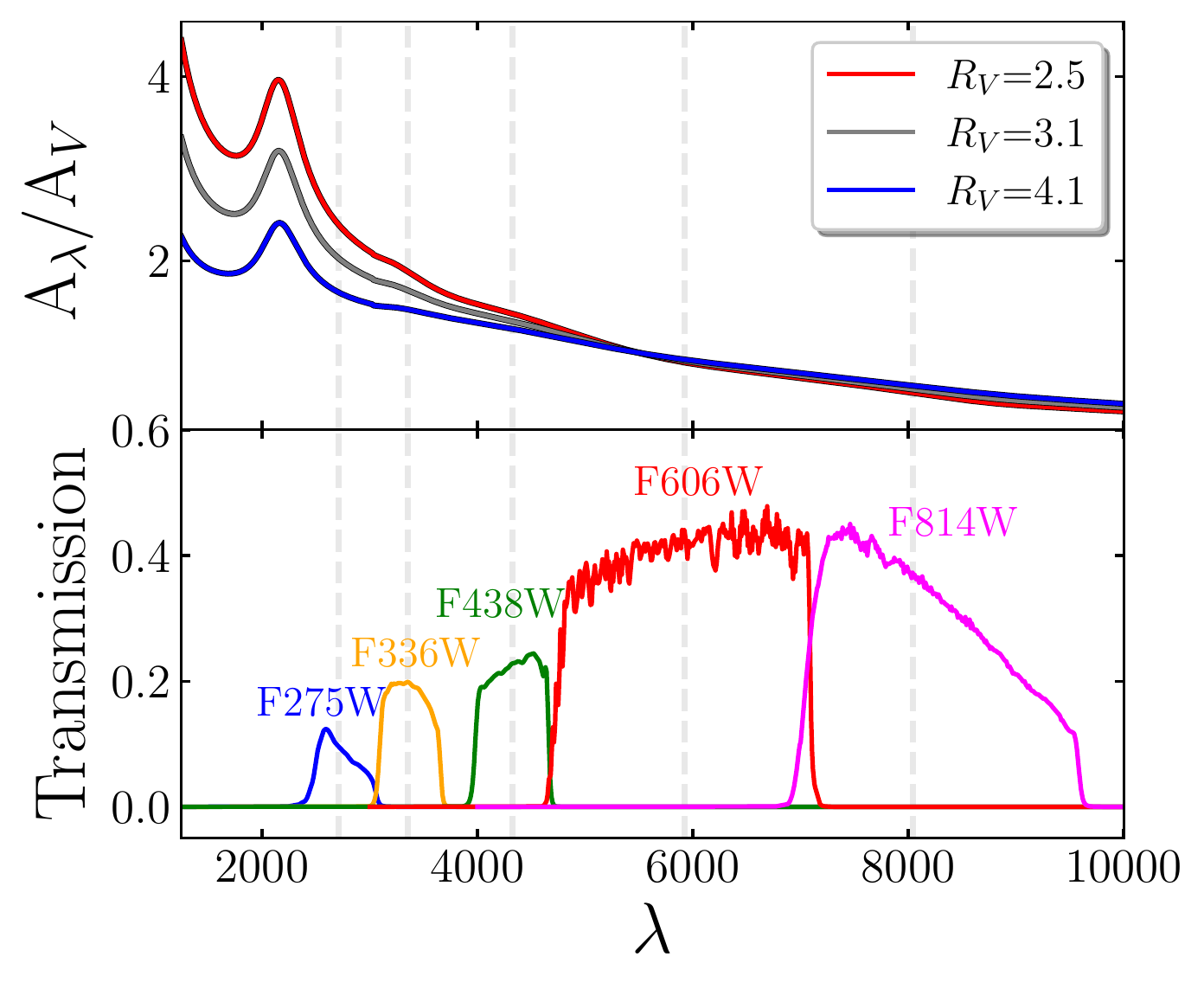}
    \caption{{\it Upper panel.} Extinction law for three different assumptions of the $R_{V}$ coefficient, namely the standard value for diffuse interstellar medium (3.1, gray line), the value commonly assumed in the direction of the Galactic Center (2.5, red line) and the value measured for the $\rho$ Ophiuchi cloud (4.1, blue line). The vertical dashed lines are drawn corresponding to the central wavelengths of the five photometric filters used in this work. {\it Bottom panel.} Transmission curves of the five filters used in this work to correct photometry for differential reddening.}
    \label{fig:ext_curves}
\end{figure}

\begin{figure} 
    \begin{center} 
    \includegraphics[width=.45\textwidth,trim={0cm 0cm 0cm 0cm},clip]{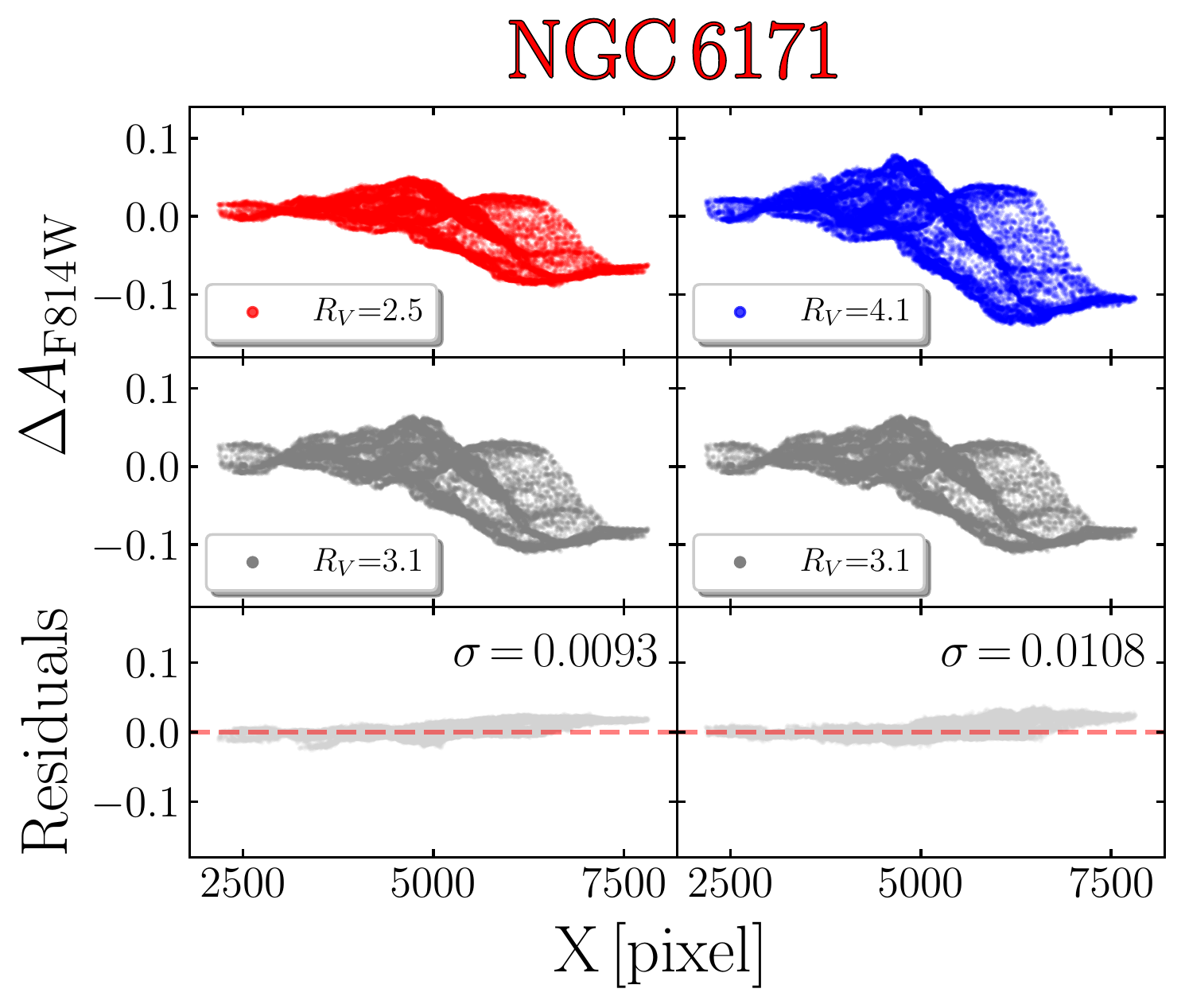}
    \caption{Local reddening variations in the direction of NGC\,6171 as a function of the X coordinates calculated by assuming $R_{V}=2.5$ (upper-left panel), $R_{V}=4.1$ (upper-right panel), and $R_{V}=3.1$ (middle panels). Bottom panels illustrate the residuals, derived by comparing $\Delta A_{F814W}$ calculated by assuming $R_{V}=2.5$ (left) and $R_{V}=4.1$ (right), respectively, with the original $\Delta A_{F814W}$, obtained by assuming $R_{V}=3.1$. The dashed red line marks the average difference, whereas the dispersion of the residuals is reported in the upper-right corner.} 
    \label{fig:respos}
    \end{center}
\end{figure} 

\begin{figure*} 
  \centering
  \includegraphics[height=10cm,trim={0cm 0cm 0cm 0cm},clip]{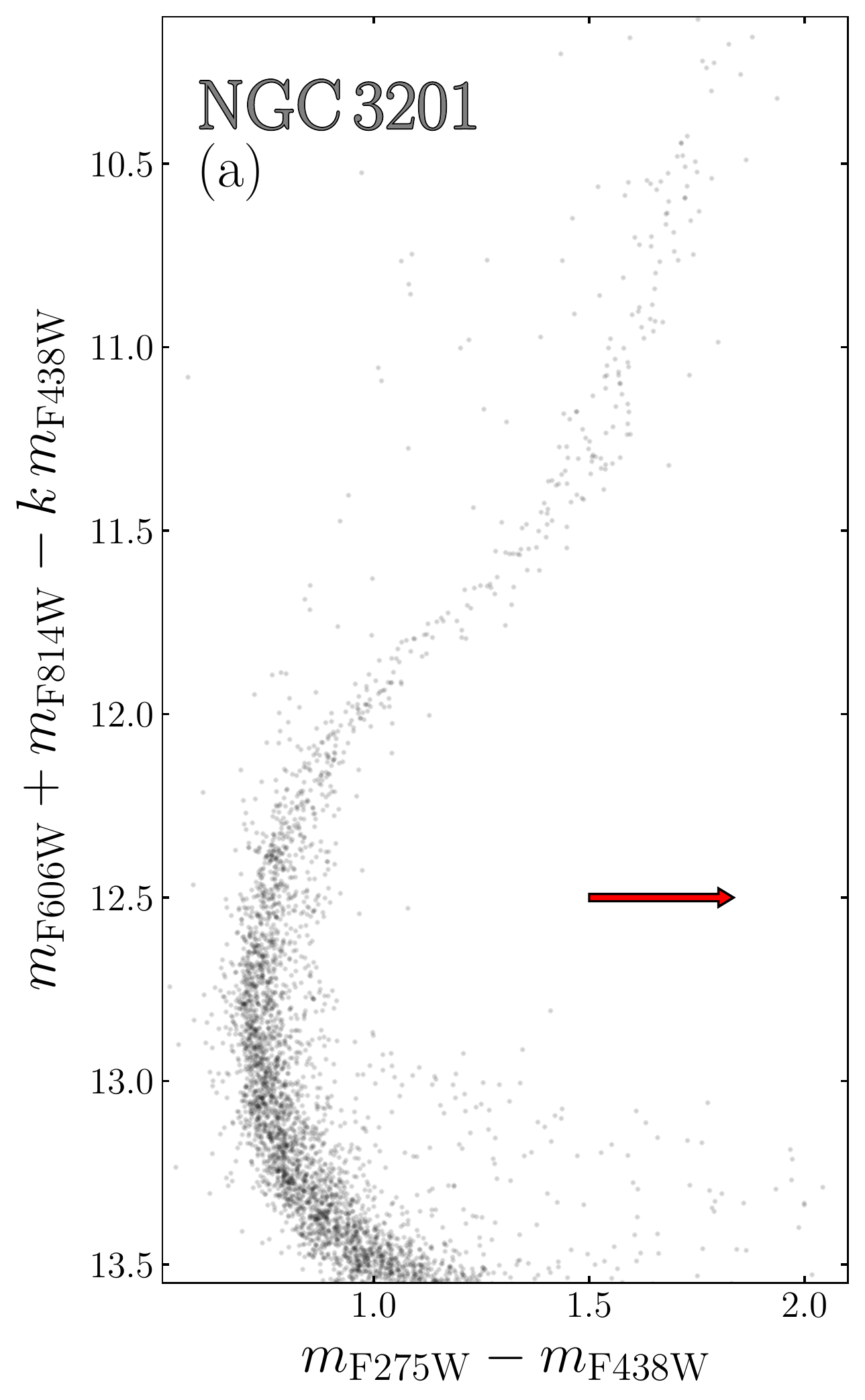}
  \includegraphics[height=10cm,trim={0cm 0cm 0cm 0cm},clip]{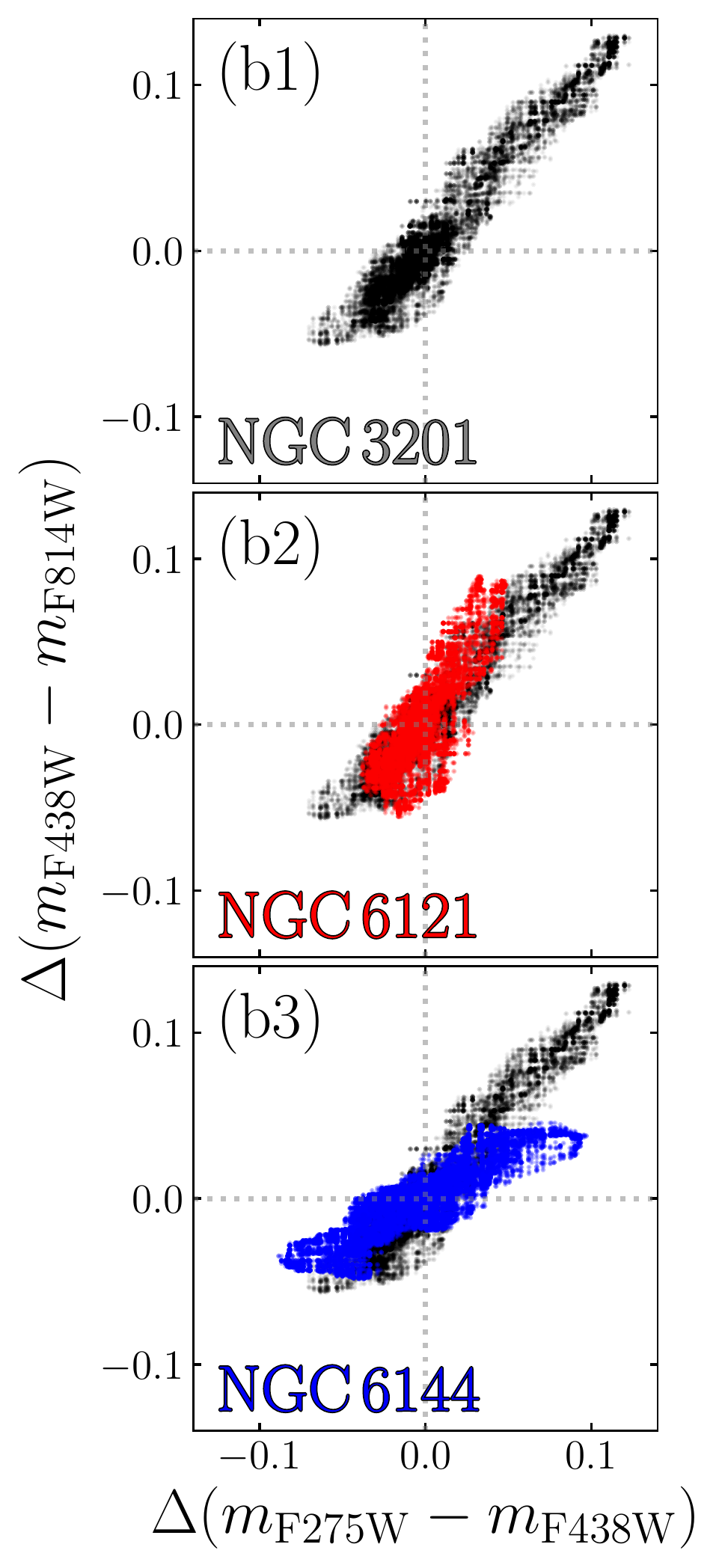}
  \caption{$m_{\rm F606W}+m_{\rm F814W}-km_{\rm F438W}$\,versus\,$m_{\rm F275W}-m_{\rm F438W}$ pseudo-CMD of NGC\,3201 (panel a). 
  $\Delta (m_{\rm F438W}-m_{\rm F814W})$\,versus\,$\Delta (m_{\rm F275W}-m_{\rm F438W})$ diagrams for NGC\,3201 (b1), NGC\,6121 (b2), and NGC\,6144 (b3).} 
  \label{fig:rlaw_method}
\end{figure*}

\begin{figure*} 
    \begin{center} 
    \includegraphics[width=.95\textwidth,trim={0cm 0cm 0cm 0cm},clip]{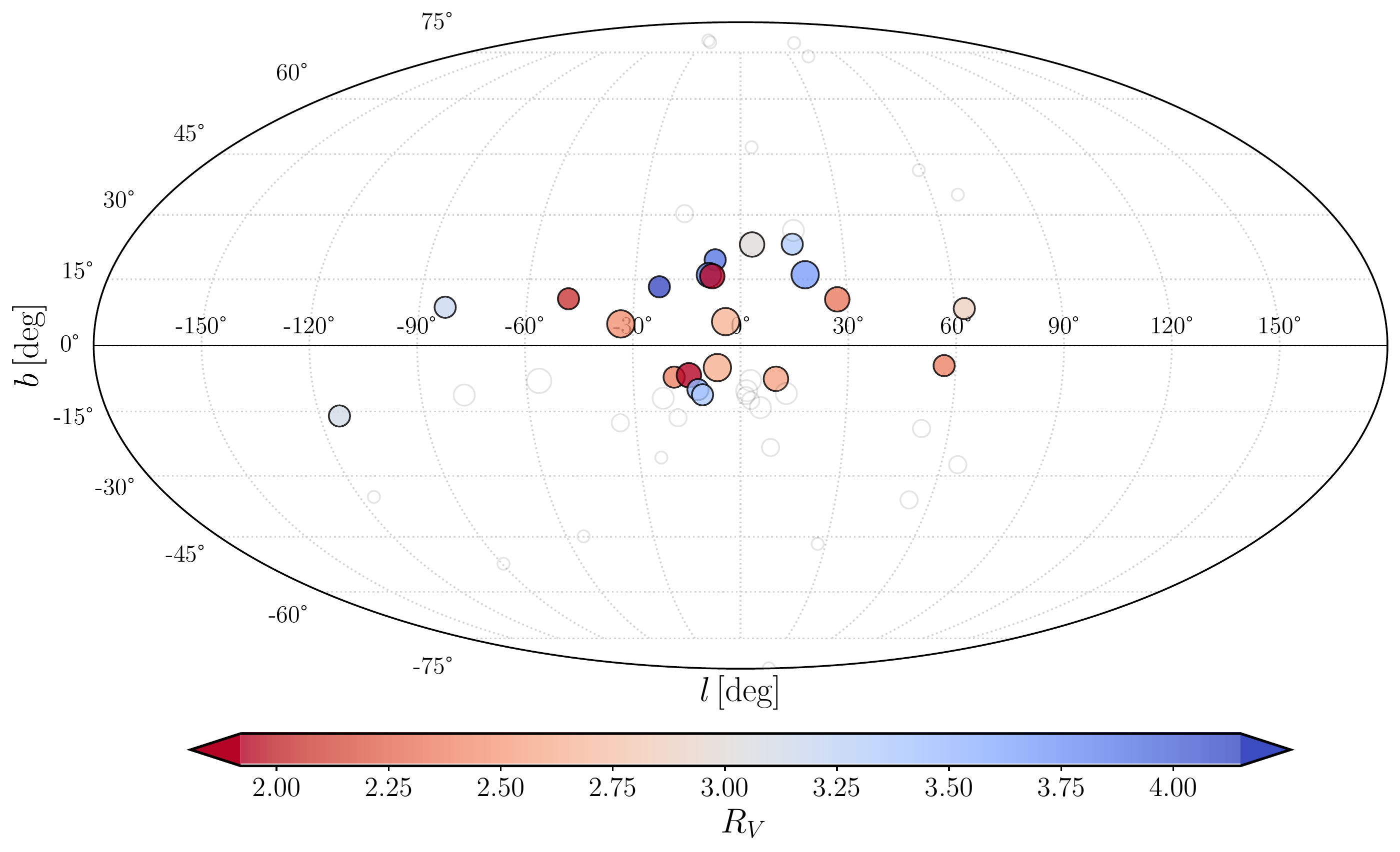}
    \caption{Spatial distribution of the 56 GCs studied in this work in Galactic coordinates. The size of each point is proportional to the average reddening of the cluster. The 21 GCs for which the corrected photometry provides improved CMDs are colour-coded according to the measured $R_{V}$ in their FoV.} 
    \label{fig:mollrv}
    \end{center}
\end{figure*} 

\begin{figure*} 
    \begin{center} 
    \includegraphics[width=.9\textwidth,trim={1.5cm 0cm 1.75cm 5.3cm},clip]{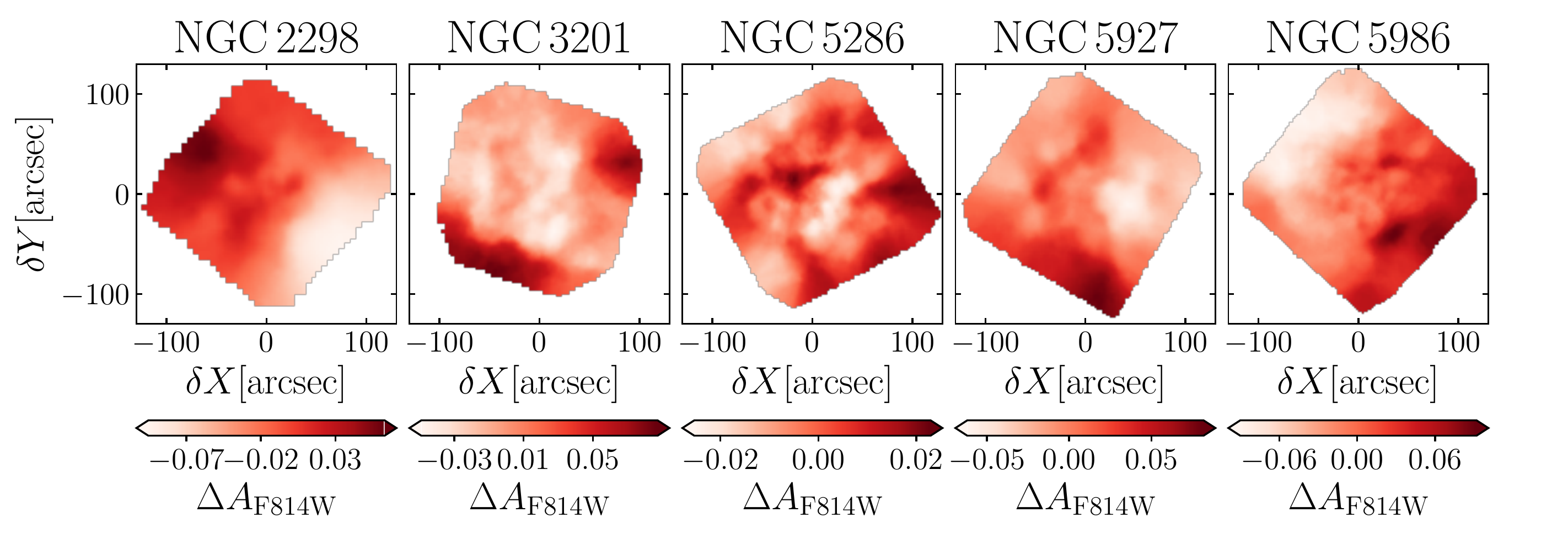}
    \includegraphics[width=.9\textwidth,trim={1.5cm 0cm 1.75cm 5.3cm},clip]{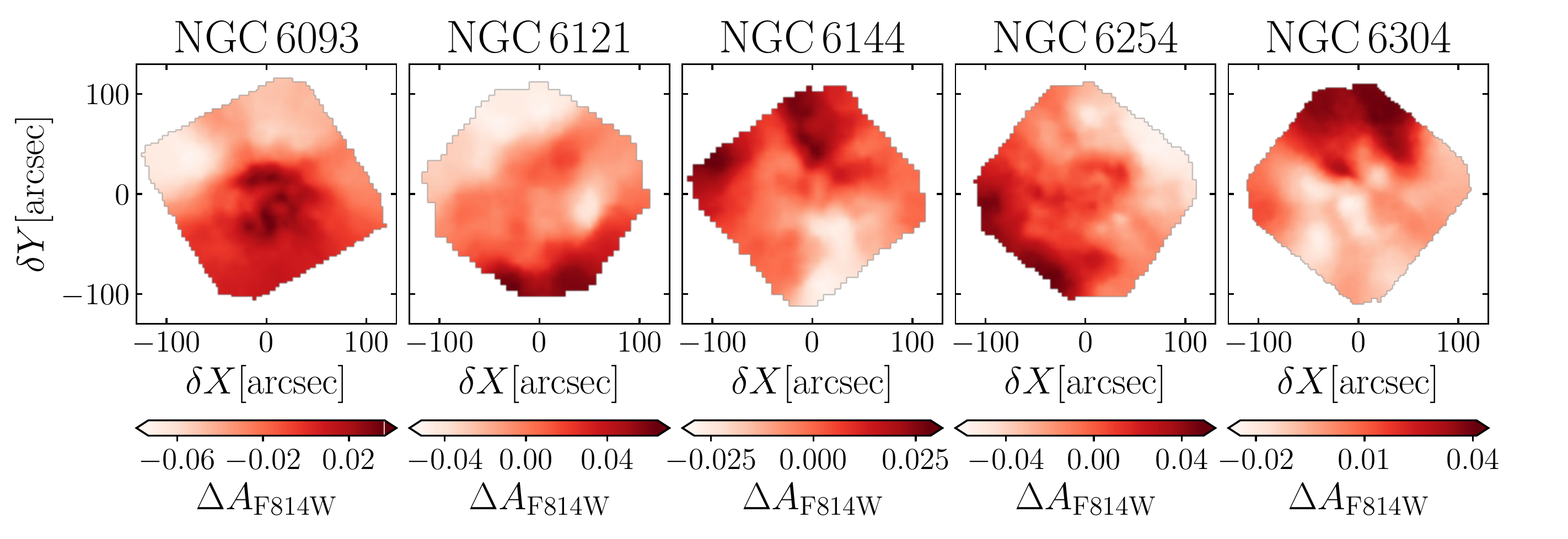}
    \includegraphics[width=.9\textwidth,trim={1.5cm 0cm 1.75cm 5.3cm},clip]{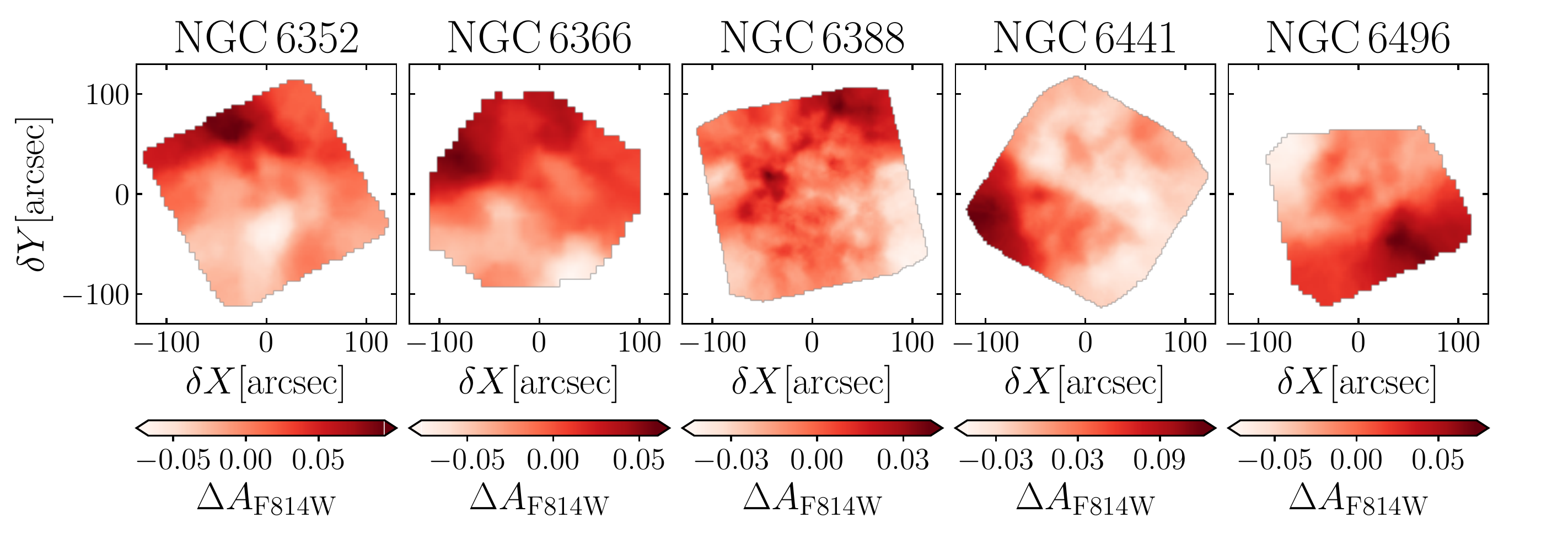}
    \includegraphics[width=.9\textwidth,trim={1.5cm 0cm 1.75cm 5.3cm},clip]{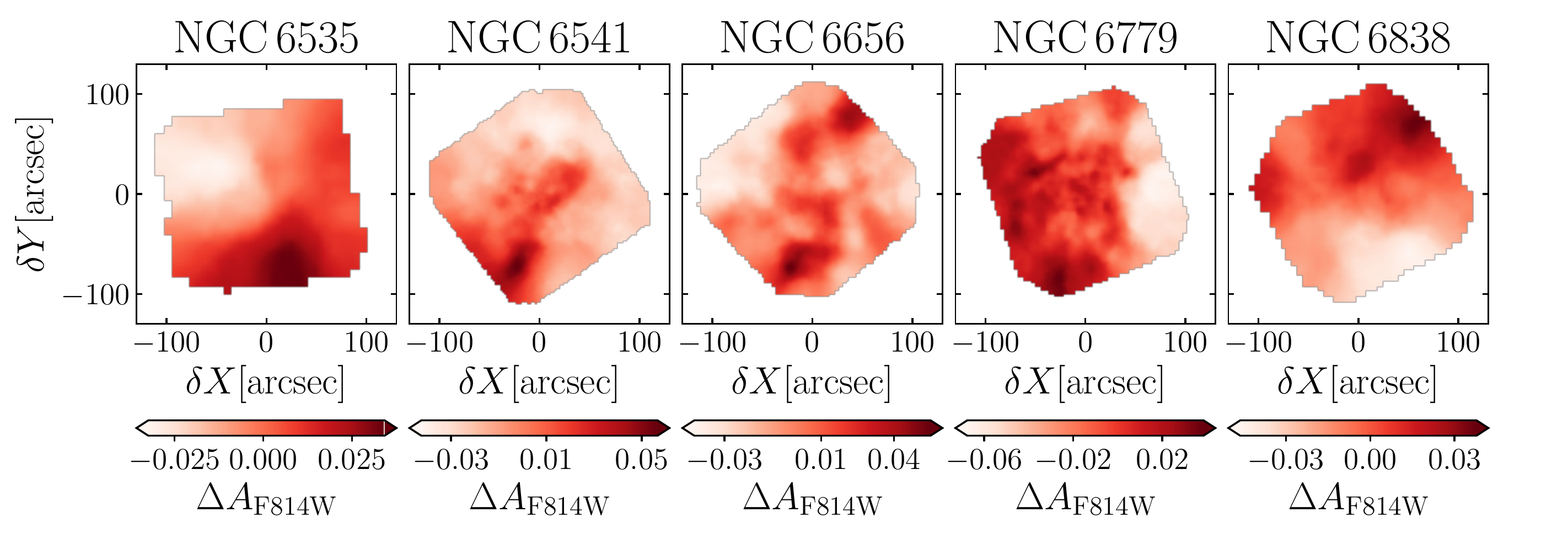}
    \caption{Differential-reddening maps in the direction of twenty Galactic GCs. The colour bar in the bottom panel of each plot codifies the $\Delta A_{\rm F814W}$ values ranging from less extinct regions (lighter colours) to more extinct areas (redder colours).} 
    \label{fig:drmaps1}
    \end{center}
\end{figure*} 

\begin{figure*} 
    \centering
    \includegraphics[width=.95\textwidth,trim={0cm 0cm 0cm 0cm},clip]{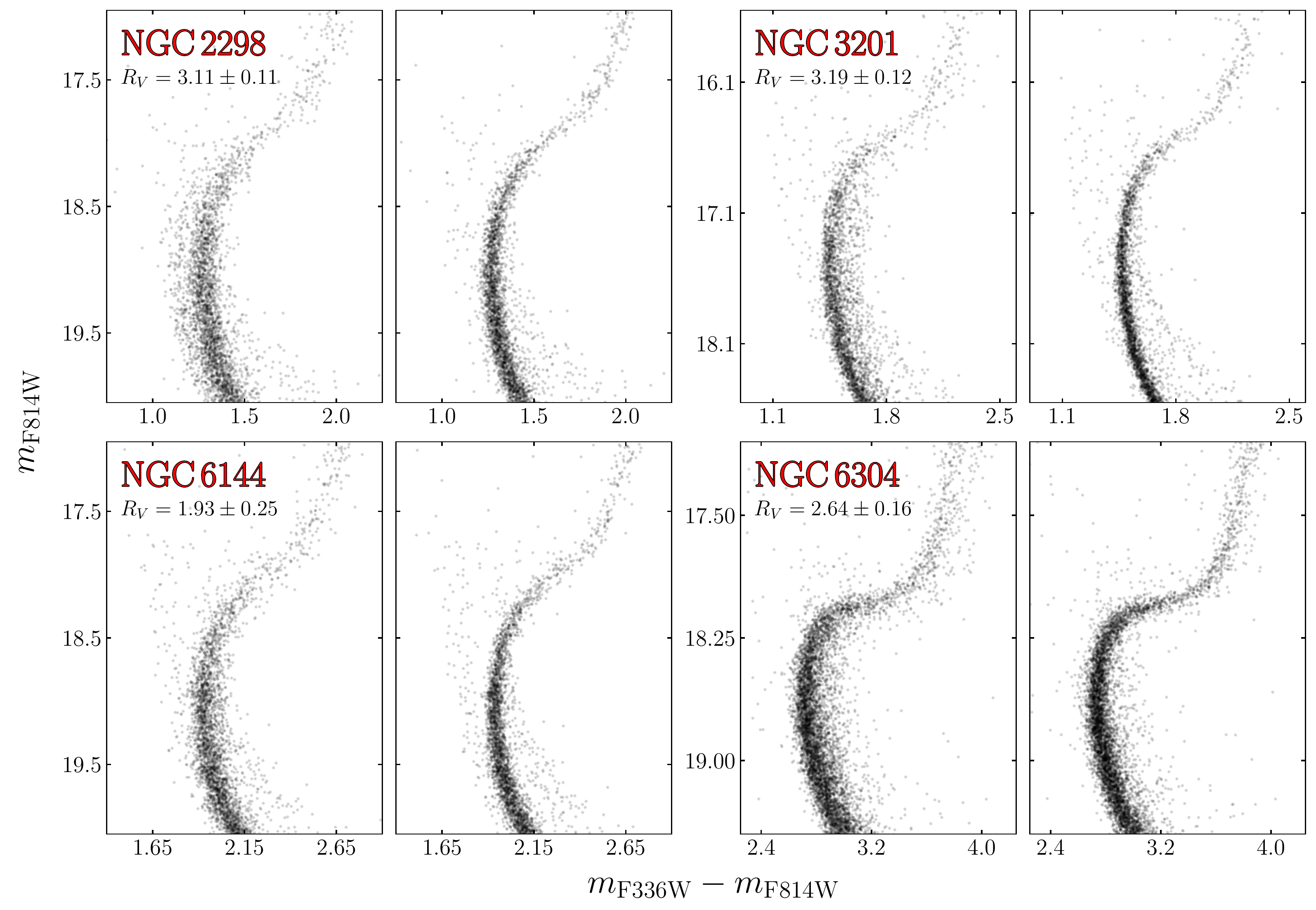}
    \caption{$m_{\rm F814W}$ versus\,$m_{\rm F336W}-m_{\rm F814W}$ CMDs for four GCs before (left panel) and after (right panel) applying the procedure to correct for differential reddening.} 
    \label{fig:cmd}
\end{figure*} 

\begin{figure*} 
    \centering
    \includegraphics[width=.95\textwidth,trim={0cm 0cm 0cm 0cm},clip]{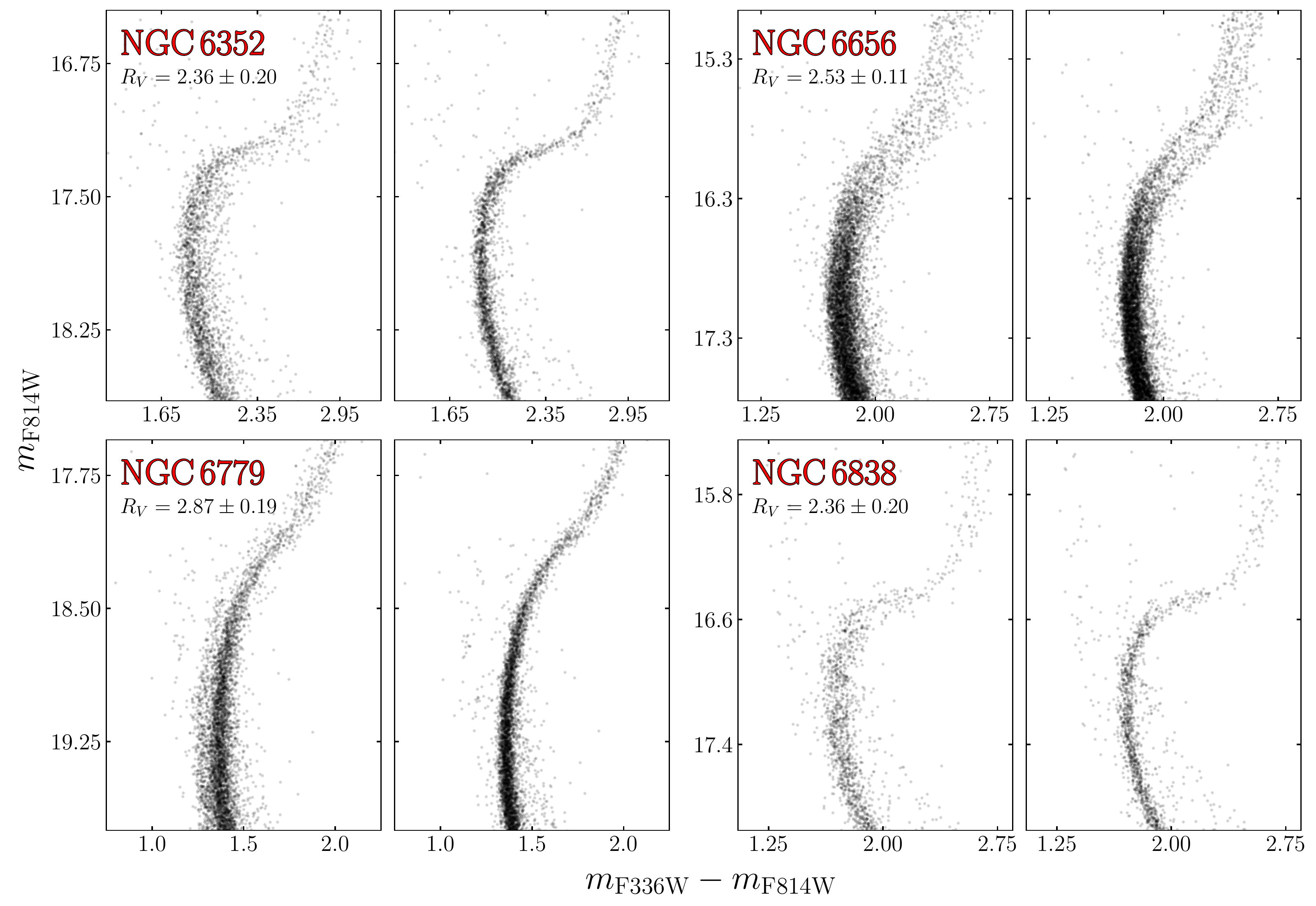}
    \caption{Same as in Fig.\,\ref{fig:cmd} but for NGC\,6352, NGC\,6656, NGC\,6779, and NGC\,6838.} 
    \label{fig:cmd2}
\end{figure*} 

\subsection{The impact of the reddening law on differential-reddening determination}
\label{redd_law}
The interstellar reddening law, namely the relation between the absorption coefficient $A_{\lambda}$ and the wavelength, has been expressed in several ways by different authors in the literature \citep[e.g.,][]{ccm89, fitzpatrick1990, odonnell1994}. All of these formulations depend on a single parameter, $R_{V}$=$A_{V}/E(B-V)$, which is usually set to 3.1, the standard value for the diffuse interstellar medium \citep[e.g.,][]{sneden1978}. However, it is well known that the extinction law is not uniform in the Galaxy, and the value of $R_{V}$ changes from place to place. For example, several authors found that $R_{V}\sim 2.5$ is an  appropriate value to describe the reddening law toward the Galactic Center \citep[e.g.,][]{udalski2003a, nataf2013, nataf2016a}. In addition, $R_{V}$ can change significantly also in nearby regions, as in the case of the $\rho$ Ophiuchi cloud, where $R_{V} = 4.1$ \citep{ccm89, hendricks2012a}. Similarly, the investigation of the various regions of recent star formation in the Large Magellanic Cloud revealed that the extinction curve is systematically flatter (in logarithmic units) than in the diffuse interstellar medium \citep{demarchi2014a, demarchi2019a, demarchi2016a, demarchi2020a, demarchi2021a}. 

In the upper panel of Fig.\,\ref{fig:ext_curves} we show three curves corresponding to the extinction laws derived assuming the \cite{odonnell1994} parametrization and three different values of $R_{V}$, namely 2.5 (red), 3.1 (gray), and 4.1 (blue). The bottom panel of Fig.\,\ref{fig:ext_curves} illustrates the transmission curves of the five filters used in this work to estimate differential reddening. 

To verify whether different values of $R_{V}$ can significantly alter the outcomes of the procedure to correct photometry, we corrected the CMDs of NGC\,6171 with the same procedure described in Section\,\ref{sec:dr} by assuming a non-universal reddening law. Specifically, we calculated differential reddening in the case of $R_{V}=2.5$ and $R_{V}=4.1$, and then we compared the results with our previous determination, derived assuming $R_{V}=3.1$. To do that, we first determined the absorption coefficients in the UVIS/WFC3 F275W, F336W, and F438W filters and in the WFC/ACS F606W and F814W bands. We used a  synthetic spectrum of a star with effective temperature $T_{\rm eff}=5900\,K$, gravity $\log g = 4.5$, and metallicity $\rm [Fe/H]=-1.5$ as reference. Then, we derived an absorbed spectrum by convolving the flux of the reference spectrum with the extinction law by \cite{odonnell1994}. To this aim, we assumed $E(B-V)=0.33$ \citep[from the 2010 version of][catalog]{harris1996} and $R_{V}=2.5$ and 4.1, respectively. The two synthetic spectra have been integrated over the transmission curves of the filters used in this paper to derive the corresponding magnitudes. Finally, by comparing the magnitudes of the absorbed and the reference spectrum we found the absorption coefficients that we used to derive the direction of the reddening vector.   

Results are illustrated in Fig.\,\ref{fig:respos}, where we show the local reddening variations as a function of the X coordinate derived assuming $R_{V}=2.5$, $R_{V}=4.1$, and $R_{V}=3.1$. The bottom panels show the residuals calculated comparing $\Delta A_{\rm F814W}$ derived assuming $R_{V}=2.5$ (left) and $R_{V}=4.1$ (right), with the original $\Delta A_{\rm F814W}$ distribution, derived assuming $R_{V}=3.1$. In both cases, we obtained an average difference close to zero and a dispersion, $\sigma \sim 0.01$ mag. We found similar results for NGC\,6441, which is the GC that shows the highest variations of reddening in the FoV. 

We conclude that changing $R_{V}$ has a moderate impact on the differential-reddening map of highly-reddened clusters. The effect is negligible for the GCs that exhibit small reddening variations in their FoV, which include most of the investigated objects. In the following Section, we constrain the reddening law in the direction of the 21 GCs that are significantly affected by differential reddening and improve their differential-reddening map. 


\section{The reddening law in the direction of 21 clusters}
\label{subsec:rl}
For 21 out of 56 GCs, the photometry corrected for differential reddening provides improved CMDs. These clusters, which are significantly affected by differential reddening, offer us the possibility to constrain the reddening law within the Galaxy in their directions. 

To do that we adopted the following iterative procedure. We first constructed the $m_{\rm F606W}+m_{\rm F814W}-k m_{\rm F438W}$ versus\,$m_{\rm F275W}-m_{\rm F438W}$ and $m_{\rm F606W}+m_{\rm F814W}-k m_{\rm F438W}$ versus\,$m_{\rm F438W}-m_{\rm F814W}$ pseudo-CMDs. The constant $k$ is chosen in such a way that the combination of magnitudes in the y-axis is reddening-free so that the differential reddening affects only the colour on the x-axis. Similar to the values of absorption in the F438W, F606W, and F814W bands, $k$ depends on the reddening law. At the first iteration, we assumed the \cite{odonnell1994} reddening law with $R_{V}=3.1$, which corresponds to $k=1.19$. As an example, we show the $m_{\rm F606W}+m_{\rm F814W}-k m_{\rm F438W}$ versus\,$m_{\rm F275W}-m_{\rm F438W}$ diagram of NGC\,3201 in panel a of Fig.\,\ref{fig:rlaw_method}, where we use a red arrow to indicate the reddening direction.

We used each pseudo-CMD to derive the amount of reddening that affects the $m_{\rm F275W}-m_{\rm F438W}$ and  $m_{\rm F438W}-m_{\rm F814W}$ of each star ($\Delta\,(m_{\rm F275W}-m_{\rm F438W})$ and $\Delta\,(m_{\rm F438W}-m_{\rm F814W})$) by using the procedure of Section\,\ref{sec:dr}.

To constrain the value of $R_{V}$ in the direction of the investigated targets we compared the slopes of the observed points in the $\Delta\,(m_{\rm F438W}-m_{\rm F814W})$ versus\,$\Delta\,(m_{\rm F275W}-m_{\rm F438W})$ diagram with the values inferred from simulated photometry. The latter comprises the same number of stars, observational errors, and $\Delta\,(m_{\rm F438W}-m_{\rm F814W})$ distribution as derived from the observed CMD. We assumed the  \cite{odonnell1994} reddening law and various values of $R_{V}$ ranging from 1.5 to 5.0 in steps of 0.1. For each value of $R_{V}$, we calculated the slope of the least-squares best-fit line in the $\Delta\,(m_{\rm F438W}-m_{\rm F814W})$ versus\,$\Delta\,(m_{\rm F275W}-m_{\rm F438W})$ plane. The value of $R_{V}$ that provides the smallest difference between the slopes of the line derived from simulated and observed points corresponds to the best determination of $R_{V}$. This ends one iteration. At this stage, we repeated the procedure above by using the updated value of $R_{V}$.

The results for NGC\,3201 ($R_{V}= 3.19\pm0.12$) are illustrated in the panel b1, whereas panels b2 and b3 compare the results obtained in the directions of NGC\,6121 (red, $R_{V}= 3.85\pm0.13$) and NGC\,6144 (blue, $R_{V}= 1.93\pm0.25$) with NGC\,3201. Clearly, the distributions of points in the $\Delta\,(m_{\rm F438W}-m_{\rm F814W})$ versus\,$\Delta\,(m_{\rm F275W}-m_{\rm F438W})$ plane are consistent with lines that have different slopes.

The resulting values of $R_{V}$ are listed in Table\,\ref{rv_new} and range from $\sim2.0$ to $\sim4.0$ with a high level of variability within the Galaxy. In Fig.\,\ref{fig:mollrv} we plot the spatial distribution of the studied GCs in Galactic coordinates, colour-coded according to the measured values of $R_{V}$. Typically, the clusters that are close to the Galactic plane and in proximity to the Galactic center exhibit smaller values of $R_{V}$ than the standard one.
  
We used the updated values of $R_{V}$ to calculate the direction of the reddening vector and then provide a more precise determination of local reddening variations in the direction of the 21 highly-reddened targets within our sample. The reddening maps in the direction of these clusters are provided in Fig.\,\ref{fig:drmaps1} (the map for NGC\,6171 is already shown in panel c of Fig.\,\ref{fig:dr_ngc6171}), whereas in Fig.\,\ref{fig:cmd} and \ref{fig:cmd2} we show the original (left panels) and corrected (right panels) $m_{\rm F814W}$ versus\,$m_{\rm F336W}-m_{\rm F814W}$ CMDs for eight GCs. These plots point out the effectiveness of the adopted method, as in all presented GCs the CMD region including the upper MS, the SGB, and the lower part of the RGB appears sharper and better defined.

For each of the 21 clusters, we publicly release the photometry of all stars in the catalogs corrected for differential reddening, together with the corresponding high-resolution reddening map. Specifically, for each star, we provide the ID \citep[from][]{nardiello2018}, the amount of differential reddening with the corresponding uncertainty, and the value of $\rho$.

\subsection{Reddening in the directions of 56 Globular Clusters}
\label{subsec:DR_lowmap}
The procedure presented in Section\,\ref{subsec:DR_correction} provides high-precision differential reddening estimates that allowed us to significantly improve the photometry of 21 GCs. However, the radii of the regions used to derive the amount of differential reddening associated with each star change from one star to another, thus making it challenging to compare the results from the different clusters.

To further quantify the amount of differential reddening and better compare the reddening variation in the direction of each cluster, we divided the FoV into a regular grid of $\sim$7\,$\times\,$7 arcsec$^{2}$ cells. We calculated the median $\Delta\,x'$ of reference stars inside each cell and converted this quantity into the corresponding differential-reddening value using the absorption coefficients. We estimated the uncertainty as the ratio between the dispersion of the differential-reddening values of the stars in each cell divided by the square root on $N-1$, where $N$ is the number of stars in that cell. We then derived the weighted average  of the four distinct estimates of $\Delta A_{\rm F814W}$ obtained from the $m_{\rm F814W}$ versus\,$m_{\rm X}-m_{\rm F814W}$ CMDs, where X=F275W, F336W, F438W, and F606W.

Table\,\ref{ebv_table} provides $\sigma_{\Delta A_{\rm F814W}}$ and $\Delta A_{\rm F814W,{\it max}}$, respectively the 68.27$^{\rm th}$ percentile and the difference between the 98$^{\rm th}$ and the 2$^{\rm nd}$ percentile of the differential-reddening distribution. We estimated the corresponding errors by means of bootstrapping statistics. Specifically, we generated 1,000 samples and for each extraction, we used the method presented above to obtain an estimate of $\sigma_{\Delta A_{\rm F814W}}$ and $\Delta A_{\rm F814W,{\it max}}$. The uncertainties associated with these two quantities correspond to the random mean scatter of the 1,000 determinations. 

Since the FoVs of all studied clusters span a similar area, the results from our analysis allow us to compare the reddening along the directions of the 56 GCs. We used $\sigma_{\Delta A_{\rm F814W}}$ and $\Delta A_{\rm F814W,{\it max}}$ to quantify the variation of extinction across the observed FoV. Specifically, we assumed $\sigma_{\Delta A_{\rm F814W}}$ and $\Delta A_{\rm F814W,{\it max}}$ as a proxy of the average and maximum differential-reddening variation, respectively.  

\begin{figure*} 
  \centering
    \includegraphics[width=.95\textwidth]{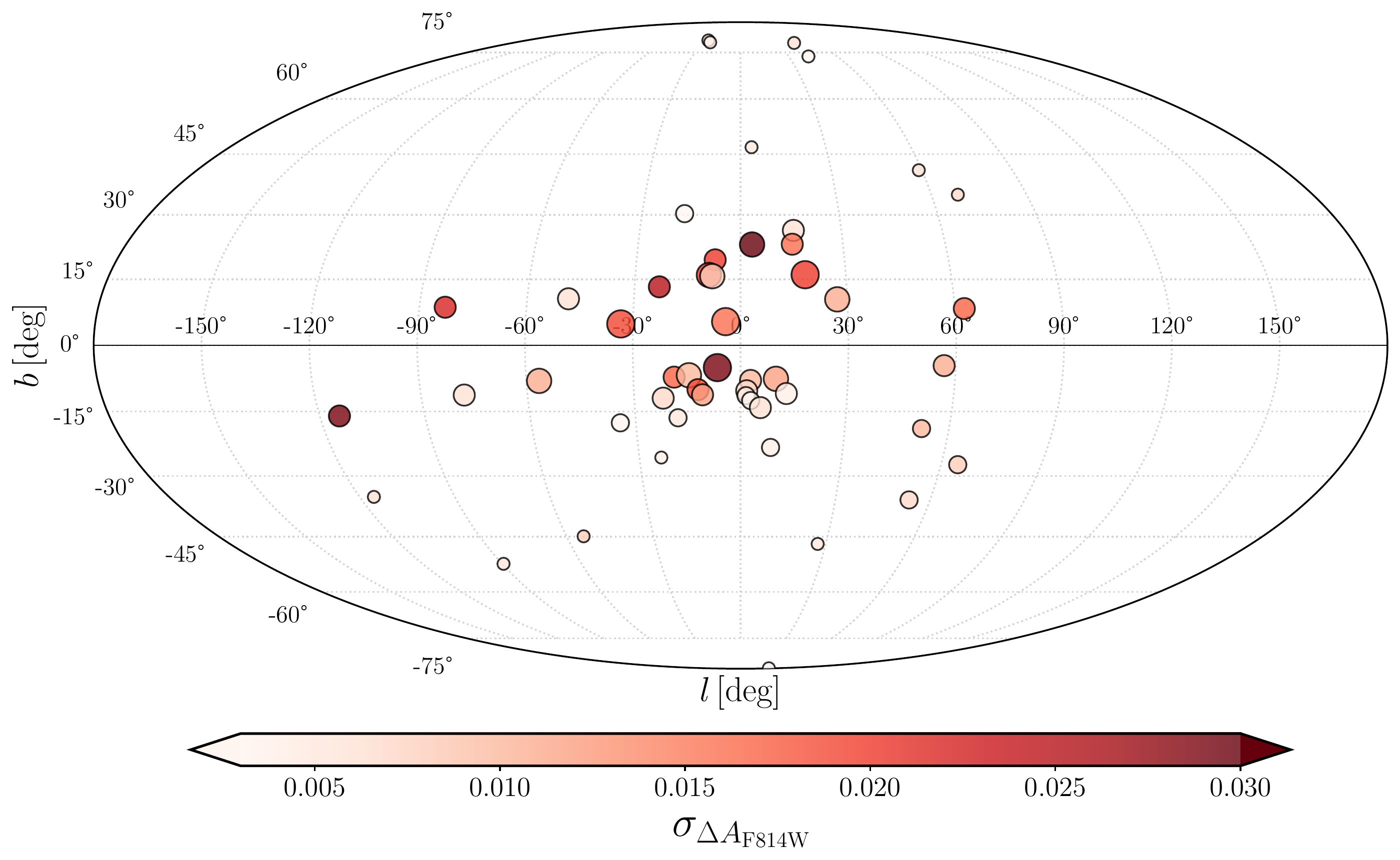}
    \caption{Spatial distribution of the 56 GCs studied in this work in Galactic coordinates. The size of each point is proportional to the average reddening of the cluster, whereas the levels of red are indicative of the 68$^{\rm th}$ percentile of the differential-reddening variation, as indicated by the colour bar on the bottom.}  
  \label{fig:redd_map}
\end{figure*}

\begin{figure*} 
  \centering
    \includegraphics[width=.95\textwidth]{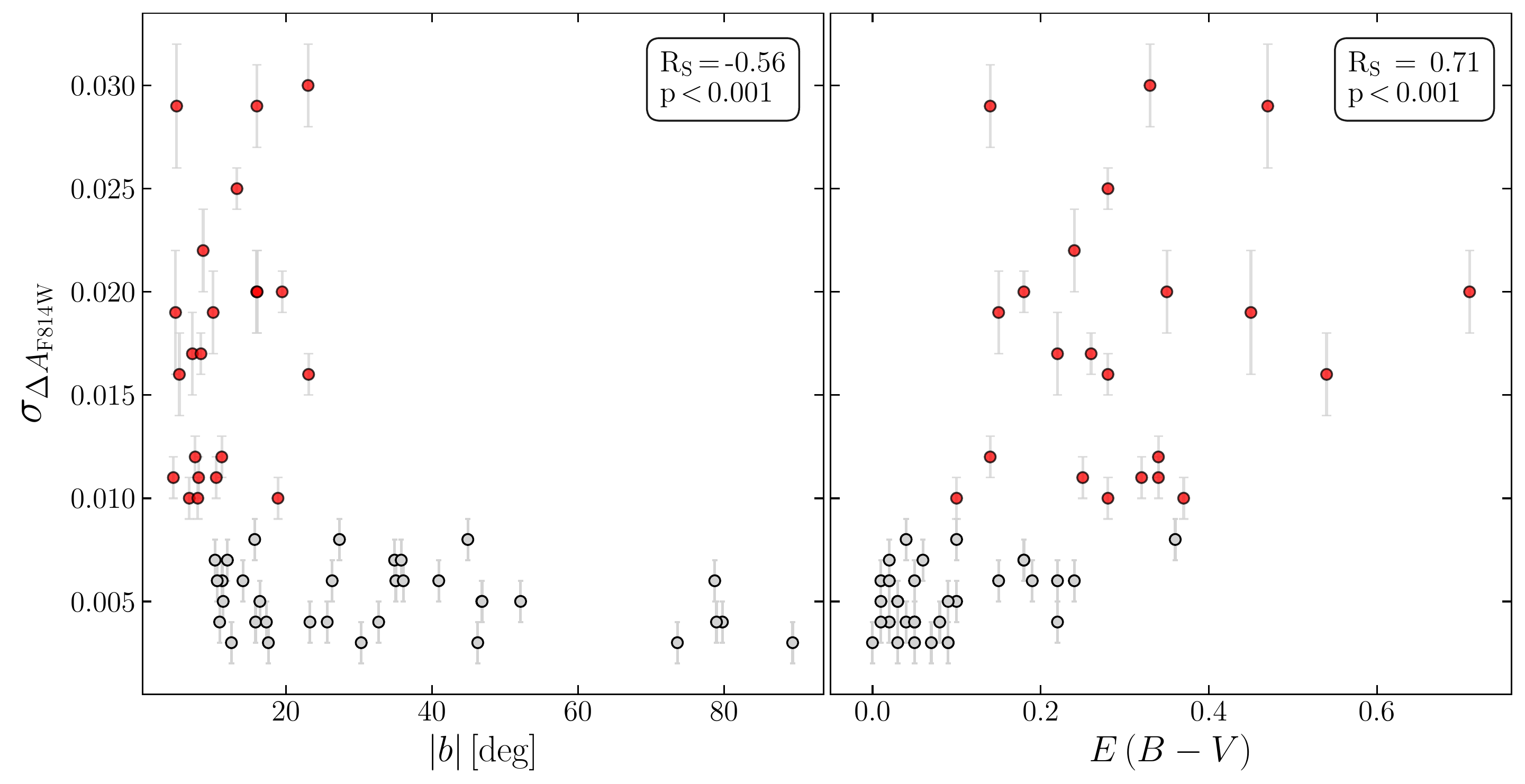}
    \caption{\textit{Left panel.} 68$^{\rm th}$ percentile of the differential-reddening variation, $\sigma_{\Delta A_{\rm F814W}}$, against the absolute value of the Galactic latitude. \textit{Right panel.} $\sigma_{\Delta A_{\rm F814W}}$ as a function of the average reddening calculated in the direction of the cluster. Red dots represent the 21 GCs for which the photometry corrected for differential reddening provides improved CMDs. The Spearman's rank correlation coefficient and the associated p-value are quoted in the upper-right corner.} 
  \label{fig:redd_sele}
\end{figure*}

The values of $\sigma_{\Delta A_{\rm F814W}}$ range from 0.003 mag to 0.030 mag, whereas $\Delta A_{\rm F814W,{\it max}}$ varies from a minimum of 0.012 mag to a maximum of 0.120 mag. As shown in Fig.\,\ref{fig:redd_map}, where we plot the spatial distribution of the studied GCs in Galactic coordinates (l,b), the GCs with large values of $\sigma_{\Delta A_{\rm F814W}}$ are located in the proximity of the Galactic plane. Specifically, $\sigma_{\Delta A_{\rm F814W}}$ exhibits a significant anti-correlation with the absolute value of the Galactic latitude (left panel of Fig.\,\ref{fig:redd_sele}), and a strong correlation with the average reddening of the host cluster \citep[from the 2010 version of the][catalog, right panel of Fig.\,\ref{fig:redd_sele}]{harris1996}. 

Our results corroborate previous findings that highly-reddened GCs typically show higher differential-reddening variations across their FoV \citep{bonatto2013, jang2022}. However, we note large variations of the $\sigma_{\Delta A_{\rm F814W}}$ value among clusters with similar reddening. A remarkable example is provided by the two clusters with the highest $E (B-V)$ values, NGC\,6366 and NGC\,6304, which are characterized by low values of $\sigma_{\Delta A_{\rm F814W}}$ with respect to the common trend. These facts would suggest that in addition to $E(B-V)$, another parameter is responsible for the value of $\sigma_{\Delta A_{\rm F814W}}$. Similar conclusions can be extended to $\Delta A_{\rm F814W,{\it max}}$.

\section{Summary and conclusions}
\label{sec:conc}
Our group has undertaken an extensive investigation of differential reddening in the directions of the Galactic and Magellanic Cloud star clusters based on multi-band photometry from {\it HST} and wide-field ground-based telescopes \citep{jang2022, milone2022a}. In the present work, we estimated the amount of differential reddening in the direction of 56 Galactic GCs based on {\it HST} photometry. To do that, we adapted the method by \cite{milone2012a} to archival data uniformly observed through the F275W, F336W, F438W of UVIS/WFC3, and the F606W, and F814W filters of WFC/ACS \citep{nardiello2018}. We derived the amount of differential reddening associated with all stars in the catalogs with a typical spatial resolution of $\sim$12 arcsec. 

We used these differential-reddening values to correct photometry for the effects of absorption. For 21 out of 56 GCs with reddening variations $\sigma_{\Delta A_{\rm F814W}}\gtrsim0.01$ we obtain improved CMDs, as all evolutionary sequences become thinner and better defined. In addition, we constrained the reddening law in the direction of these clusters. To do that, we compared the slopes of the observed points in the $\Delta\,(m_{\rm F275W}-m_{\rm F438W})$\,versus\,$\Delta\,(m_{\rm F438W}-m_{\rm F814W})$ diagram with the values inferred from simulated photometry derived assuming various values of $R_{V}$. We found that $R_{V}$ exhibits a high level of variability within the Galaxy, with values ranging from $\sim2.0$ to $\sim4.0$. The updated values of $R_{V}$ have been used to calculate the direction of the reddening vector and then derive a more precise determination of local reddening variations within the FoV of the 21 highly-reddened targets. For these clusters, we publicly release the photometric catalogs corrected for differential reddening, together with the corresponding reddening map. 

To quantify the variation of reddening across the observed FoV and compare the results from the different clusters, we computed the 68.27$^{\rm th}$ percentile ($\sigma_{\Delta A_{\rm F814W}}$) and the difference between the 98$^{\rm th}$ and the 2$^{\rm nd}$ percentile ($\Delta A_{\rm F814W,{\it max}}$) of the differential-reddening distribution. We found that $\sigma_{\Delta A_{\rm F814W}}$ ranges from 0.003 to 0.030 mag, whereas $\Delta A_{\rm F814W,{\it max}}$ varies from 0.012 to 0.120 mag. The 68.27$^{\rm th}$ percentile of the differential-reddening distribution exhibits a strong correlation with the average reddening of the host cluster. In addition, GCs with large values of $\sigma_{\Delta A_{\rm F814W}}$ are located in the proximity of the Galactic plane, as we found that there is a significant anti-correlation between $\sigma_{\Delta A_{\rm F814W}}$ and the absolute module of the Galactic latitude. Similar results have been found for $\Delta A_{\rm F814W,{\it max}}$, thus corroborating the conclusion that GCs with high values of average reddening are affected by large amounts of differential reddening.

\section*{Acknowledgements}
We thank the anonymous referee for various suggestions that improved the quality of the manuscript. This work has received support from the European Research Council (ERC) under the European Union's Horizon 2020 research innovation programme (Grant Agreement ERC-StG 2016, No 716082 'GALFOR', PI: Milone, http://progetti.dfa.unipd.it/GALFOR).
APM and ED acknowledge support from MIUR through the FARE project R164RM93XW SEMPLICE (PI: Milone) and from MIUR under PRIN program 2017Z2HSMF (PI: Bedin). 

\section*{Data Availability}
The data underlying this article will be shared on reasonable request to the corresponding author.

\begin{table*}
    \centering
    \caption{This table lists for each cluster the average reddening \citep[from the 2010 version of the][catalog]{harris1996}, the 68$^{\rm th}$ percentile of the differential-reddening distribution, the maximum variation of differential reddening across the FoV of the cluster (difference between the 98$^{\rm th}$ and the 2$^{\rm nd}$ percentile of the differential-reddening distribution), the minimum, maximum, and median radius of the regions used to estimate differential reddening in the observed FoV.} 
    \setlength{\tabcolsep}{20pt} 
    \renewcommand{\arraystretch}{1.} 
    \begin{tabular}{cccccccc}
    \hline \\[-.3cm]
    Cluster ID & $E(B-V)$ & $\sigma_{\Delta A_{\rm F814W}}$ & $\Delta A_{\rm F814W,{\it max}}$ & $\rho_{min}$ & $\rho_{max}$ & $\rho_{med}$\\
    & (mag) & (mag) & (mag) & (arcsec) & (arcsec) & (arcsec) \\[.1cm]
    \hline \\[-.4cm]
    NGC\,0104 & 0.04 & $0.008\pm0.001$ & $0.031\pm0.004$ &  4.1 & 66.2 &  8.4 \\	     
    NGC\,0288 & 0.03 & $0.003\pm0.001$ & $0.012\pm0.003$ &  9.1 & 39.1 & 13.6 \\	    
    NGC\,0362 & 0.05 & $0.003\pm0.001$ & $0.013\pm0.001$ &  4.1 & 48.9 &  6.9 \\	    
    NGC\,1261 & 0.01 & $0.005\pm0.001$ & $0.016\pm0.002$ &  3.8 & 40.1 &  9.0 \\	    
    NGC\,1851 & 0.02 & $0.006\pm0.001$ & $0.022\pm0.002$ &  3.8 & 32.9 &  7.2 \\	    
    NGC\,2298 & 0.14 & $0.029\pm0.002$ & $0.098\pm0.008$ &  5.2 & 57.6 & 16.2 \\	    
    NGC\,2808 & 0.22 & $0.006\pm0.001$ & $0.021\pm0.001$ &  5.5 & 23.8 &  8.0 \\	    
    NGC\,3201 & 0.24 & $0.022\pm0.002$ & $0.078\pm0.005$ &  6.6 & 32.2 & 12.7 \\	    
    NGC\,4590 & 0.05 & $0.006\pm0.001$ & $0.020\pm0.002$ &  4.1 & 38.6 & 10.6 \\	    
    NGC\,4833 & 0.32 & $0.011\pm0.001$ & $0.044\pm0.006$ &  4.3 & 21.0 &  8.1 \\	    
    NGC\,5024 & 0.02 & $0.004\pm0.001$ & $0.014\pm0.002$ &  4.8 & 35.7 &  8.2 \\	    
    NGC\,5053 & 0.01 & $0.004\pm0.001$ & $0.015\pm0.002$ & 12.9 & 50.3 & 20.2 \\	    
    NGC\,5272 & 0.01 & $0.006\pm0.001$ & $0.020\pm0.001$ &  4.8 & 56.6 &  8.0 \\	    
    NGC\,5286 & 0.24 & $0.006\pm0.001$ & $0.022\pm0.004$ &  5.2 & 40.5 &  9.5 \\	    
    NGC\,5466 & 0.00 & $0.003\pm0.001$ & $0.013\pm0.002$ &  9.8 & 39.1 & 15.3 \\	    
    NGC\,5897 & 0.09 & $0.003\pm0.001$ & $0.012\pm0.001$ &  7.5 & 30.4 & 12.2 \\	    
    NGC\,5904 & 0.03 & $0.005\pm0.001$ & $0.016\pm0.001$ &  4.0 & 71.4 &  9.0 \\	     
    NGC\,5927 & 0.45 & $0.019\pm0.003$ & $0.076\pm0.012$ &  5.7 & 43.3 & 12.5 \\	    
    NGC\,5986 & 0.28 & $0.025\pm0.001$ & $0.086\pm0.004$ &  3.9 & 37.1 &  8.0 \\	    
    NGC\,6093 & 0.18 & $0.020\pm0.001$ & $0.066\pm0.005$ &  5.4 & 52.1 & 10.1 \\	    
    NGC\,6101 & 0.05 & $0.004\pm0.001$ & $0.014\pm0.001$ &  5.7 & 36.4 & 11.5 \\	    
    NGC\,6121 & 0.35 & $0.020\pm0.002$ & $0.082\pm0.012$ & 11.3 & 43.1 & 19.8 \\	    
    NGC\,6144 & 0.36 & $0.008\pm0.001$ & $0.028\pm0.002$ &  6.4 & 32.7 & 11.7 \\	    
    NGC\,6171 & 0.33 & $0.030\pm0.002$ & $0.100\pm0.004$ &  5.5 & 39.2 & 11.9 \\	    
    NGC\,6205 & 0.02 & $0.006\pm0.001$ & $0.021\pm0.001$ &  4.3 & 35.8 &  7.5 \\	    
    NGC\,6218 & 0.19 & $0.006\pm0.001$ & $0.022\pm0.001$ &  6.6 & 28.8 & 12.1 \\	    
    NGC\,6254 & 0.28 & $0.016\pm0.001$ & $0.052\pm0.003$ &  4.1 & 27.2 &  8.7 \\	    
    NGC\,6304 & 0.54 & $0.016\pm0.002$ & $0.061\pm0.007$ &  5.5 & 47.7 & 18.3 \\	    
    NGC\,6341 & 0.02 & $0.007\pm0.001$ & $0.022\pm0.001$ &  4.2 & 30.5 &  8.5 \\	    
    NGC\,6352 & 0.22 & $0.017\pm0.002$ & $0.070\pm0.010$ &  7.3 & 40.2 & 14.4 \\	    
    NGC\,6362 & 0.09 & $0.003\pm0.001$ & $0.013\pm0.001$ &  7.3 & 36.3 & 12.3 \\	    
    NGC\,6366 & 0.71 & $0.020\pm0.002$ & $0.078\pm0.006$ & 12.4 & 46.0 & 20.8 \\	    
    NGC\,6388 & 0.37 & $0.010\pm0.001$ & $0.038\pm0.005$ &  5.5 & 39.4 &  8.1 \\	    
    NGC\,6397 & 0.18 & $0.007\pm0.001$ & $0.025\pm0.003$ &  3.1 & 40.6 & 13.2 \\	    
    NGC\,6441 & 0.47 & $0.029\pm0.003$ & $0.120\pm0.013$ &  6.2 & 39.6 &  9.1 \\	    
    NGC\,6496 & 0.15 & $0.019\pm0.002$ & $0.072\pm0.007$ &  5.6 & 43.6 & 13.0 \\	    
    NGC\,6535 & 0.34 & $0.011\pm0.001$ & $0.043\pm0.006$ &  6.6 & 81.9 & 25.9 \\	    
    NGC\,6541 & 0.14 & $0.012\pm0.001$ & $0.048\pm0.006$ &  4.5 & 34.9 & 10.4 \\	    
    NGC\,6584 & 0.10 & $0.005\pm0.001$ & $0.019\pm0.002$ &  3.1 & 41.2 &  8.1 \\	    
    NGC\,6624 & 0.28 & $0.010\pm0.001$ & $0.042\pm0.008$ &  3.8 & 43.5 & 11.7 \\	    
    NGC\,6637 & 0.18 & $0.007\pm0.001$ & $0.025\pm0.002$ &  2.8 & 34.7 &  7.3 \\	    
    NGC\,6652 & 0.09 & $0.005\pm0.001$ & $0.020\pm0.003$ &  3.0 & 50.8 & 12.9 \\	    
    NGC\,6656 & 0.34 & $0.012\pm0.001$ & $0.047\pm0.002$ &  7.3 & 30.9 & 12.4 \\	    
    NGC\,6681 & 0.07 & $0.003\pm0.001$ & $0.013\pm0.001$ &  2.9 & 37.4 &  9.9 \\	    
    NGC\,6715 & 0.15 & $0.006\pm0.001$ & $0.022\pm0.002$ &  6.2 & 70.0 & 11.2 \\	    
    NGC\,6717 & 0.22 & $0.004\pm0.001$ & $0.017\pm0.003$ &  4.6 & 54.4 & 23.3 \\	    
    NGC\,6723 & 0.05 & $0.004\pm0.001$ & $0.017\pm0.002$ &  3.4 & 24.2 &  7.2 \\	    
    NGC\,6752 & 0.04 & $0.004\pm0.001$ & $0.015\pm0.002$ &  2.6 & 85.7 & 10.0 \\	    
    NGC\,6779 & 0.26 & $0.017\pm0.001$ & $0.057\pm0.003$ &  2.9 & 29.4 &  7.6 \\	    
    NGC\,6809 & 0.08 & $0.004\pm0.001$ & $0.014\pm0.001$ &  8.6 & 31.9 & 14.0 \\	    
    NGC\,6838 & 0.25 & $0.011\pm0.001$ & $0.039\pm0.002$ & 11.9 & 44.6 & 19.2 \\	    
    NGC\,6934 & 0.10 & $0.010\pm0.001$ & $0.034\pm0.005$ &  4.1 & 44.5 &  9.4 \\	    
    NGC\,6981 & 0.05 & $0.004\pm0.001$ & $0.015\pm0.001$ &  3.6 & 42.1 &  9.5 \\	    
    NGC\,7078 & 0.10 & $0.008\pm0.001$ & $0.029\pm0.002$ &  5.2 & 33.3 &  8.3 \\	    
    NGC\,7089 & 0.06 & $0.007\pm0.001$ & $0.025\pm0.001$ &  5.5 & 34.7 &  8.3 \\	    
    NGC\,7099 & 0.03 & $0.005\pm0.001$ & $0.018\pm0.003$ &  2.4 & 29.8 &  8.2 \\
\hline \hline
\end{tabular}
\label{ebv_table}
\end{table*}
\begin{table}
    \centering    
    \setlength{\tabcolsep}{10pt} 
    \renewcommand{\arraystretch}{1.} 
    \begin{tabular}{cccc}
    \hline \\[-.3cm]
    Cluster ID & $R_{V}$ & Cluster ID & $R_{V}$  \\[.1cm]
    \hline \\[-.4cm]
    NGC\,2298 & $3.11\pm0.11$ & NGC\,6352 & $2.36\pm0.20$ \\	        
    NGC\,3201 & $3.19\pm0.12$ & NGC\,6366 & $3.70\pm0.43$ \\	    	    
    NGC\,5286 & $2.04\pm0.28$ & NGC\,6388 & $1.92\pm0.24$ \\	    	    
    NGC\,5927 & $2.43\pm0.16$ & NGC\,6441 & $2.60\pm0.43$ \\	    	    
    NGC\,5986 & $4.15\pm0.44$ & NGC\,6496 & $3.53\pm0.23$ \\	    	    
    NGC\,6093 & $3.93\pm0.18$ & NGC\,6535 & $2.32\pm0.18$ \\	    	    	    
    NGC\,6121 & $3.85\pm0.13$ & NGC\,6541 & $3.40\pm0.19$ \\	    	    
    NGC\,6144 & $1.93\pm0.25$ & NGC\,6656 & $2.53\pm0.11$ \\	    	    
    NGC\,6171 & $3.01\pm0.11$ & NGC\,6779 & $2.87\pm0.19$ \\	    	    
    NGC\,6254 & $3.35\pm0.27$ & NGC\,6838 & $2.36\pm0.20$ \\	    	    
    NGC\,6304 & $2.64\pm0.16$ & & \\	
    \hline \hline
    \end{tabular}
    \caption{This table lists for each of the 21 highly-reddened GCs the new values of $R_{V}$ with the corresponding error calculated in Section\,\ref{subsec:rl}.}
    \label{rv_new}
\end{table}


\bibliography{example}{} 

\appendix

\section{Multiple populations and differential reddening}
\label{MP_DR}
All GCs studied in this work show multiple stellar populations with different chemical compositions that define distinct sequences along the entire CMD \citep{milone2017a}. In this appendix, we investigate whether or not the presence of multiple populations in a GC affects the differential reddening determination, described in Section\,\ref{sec:dr}. 
  
\begin{figure}
    \centering    
    \includegraphics[width=.45\textwidth,trim={0cm 0cm 0cm 0cm},clip]{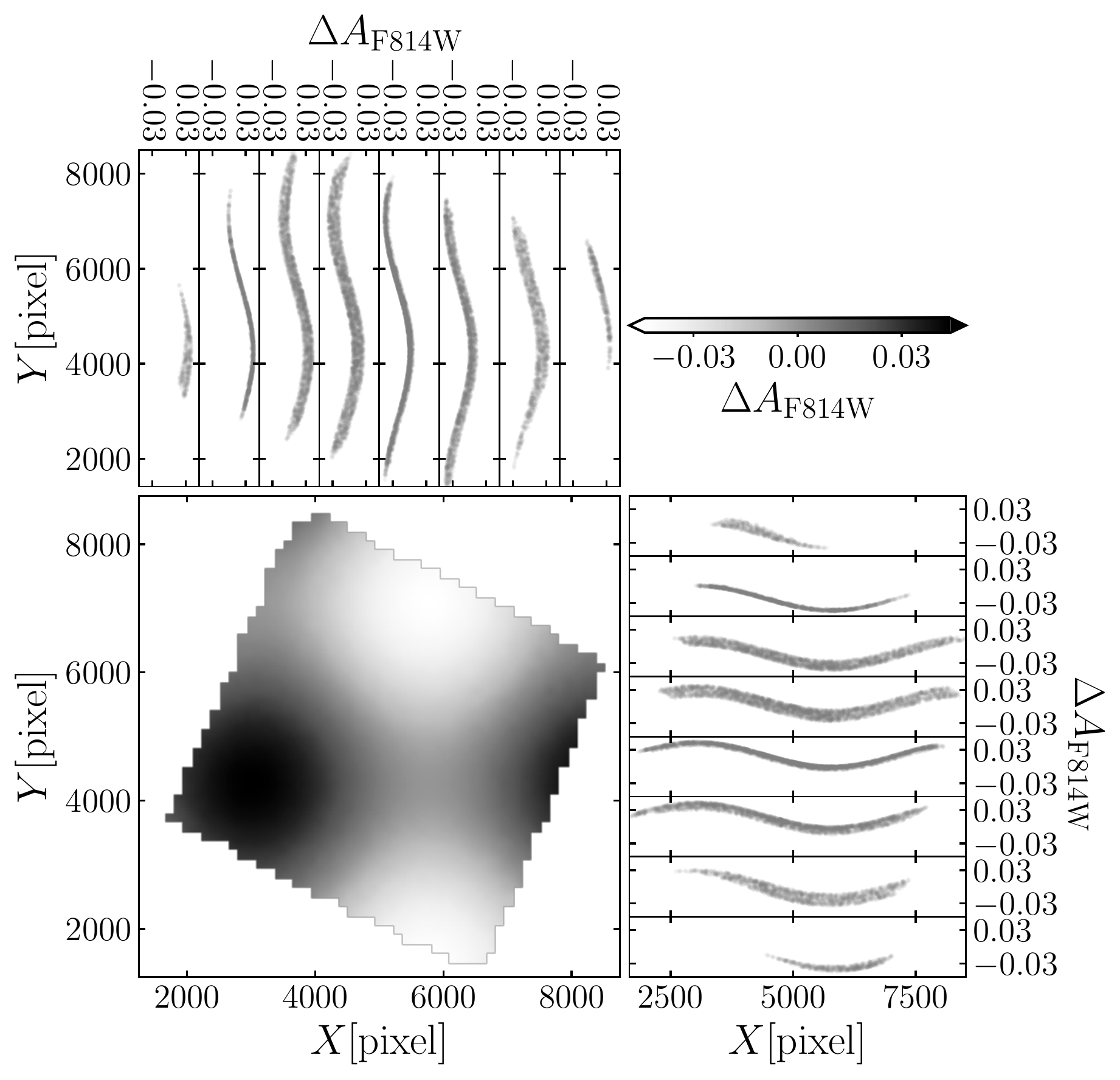}
    \caption{Differential-reddening map added to the simulated CMDs. The top and right panels show $\Delta A_{\rm F814W}$ as a function of the X and Y coordinate of the horizontal and vertical slices in which we divided the FoV.} 
    \label{fig:rmap_inp}
\end{figure}

To do that, we first simulated the $m_{\rm F814W}$ versus\,$m_{\rm X}-m_{\rm F814W}$ CMDs, where X=F275W, F336W, F438W, and F606W, to which we have added the map of differential reddening shown in the left panel of Figure\,\ref{fig:rmap_inp}. In close analogy with \citet[][see their appendix A]{milone2012a}, the reddening variations that we simulated are related to stellar positions (X, Y) by the following relations:
\begin{equation}
    \Delta\,E(B-V)=C_{1}(\cos{X'}+\sin{Y'})
\end{equation}
where 
\begin{align*}
    X' = C_{2}\pi(X-X_{MAX})/(X_{MAX}-X_{MIN}),
\end{align*}
\begin{align*}
    Y' = C_{2}\pi(Y-Y_{MAX})/(Y_{MAX}-Y_{MIN}).
\end{align*}
Specifically, $X_{MIN,MAX}$ and $Y_{MIN,MAX}$ correspond to the minimum and maximum values of X and Y, respectively, whereas $C_{1}$ and $C_{2}$ are two free parameters determining how the reddening variations are distributed across the FoV of the cluster (see \citealt{milone2012a} for details). We fixed $C_{1}=0.012$ and $C_{2}=2.5$ in such a way that the simulated map resembles the one of NGC\,6838. This cluster, which has been widely studied in the context of multiple populations, hosts first and second-population stars with different chemical compositions. The stellar populations of NGC\,6838 can be followed continuously along various evolutionary sequences, including the red horizontal branch, the RGB, the SGB, and the MS \citep{cordoni2020a, dondoglio2021a, legnardi2022}. 

To investigate the impact of multiple populations on the determination of differential reddening, we simulated two groups of CMDs. We used the same number of stars and the same F814W luminosity distribution as observed in NGC\,6838. We first derived CMDs where stars belong to a single population. We assumed for all stars the colours and magnitudes inferred from first-population stars by \citet{legnardi2022}. Then, we simulated CMDs composed of first- and second-population stars. In this case, second-population stars comprise 70$\%$ of the total number of stars, which is a typical value for Galactic GCs \citep{milone2017a, dondoglio2021a}. We adopted the colours and magnitudes of the two stellar populations measured by \citet{legnardi2022}.

We corrected these CMDs with the same procedure described in Section\,\ref{sec:dr} obtaining a reddening map that is similar to the one that we have simulated. We assumed the differences between the measured differential-reddening values ($\Delta A_{\rm F814W}^{\rm OUT}$) and the simulated ones ($\Delta A_{\rm F814W}^{\rm IN}$) as an estimate of the effectiveness of the correction procedure.

\begin{figure}
    \centering
    \includegraphics[width=.45\textwidth,trim={0cm 0cm 0cm 0cm},clip]{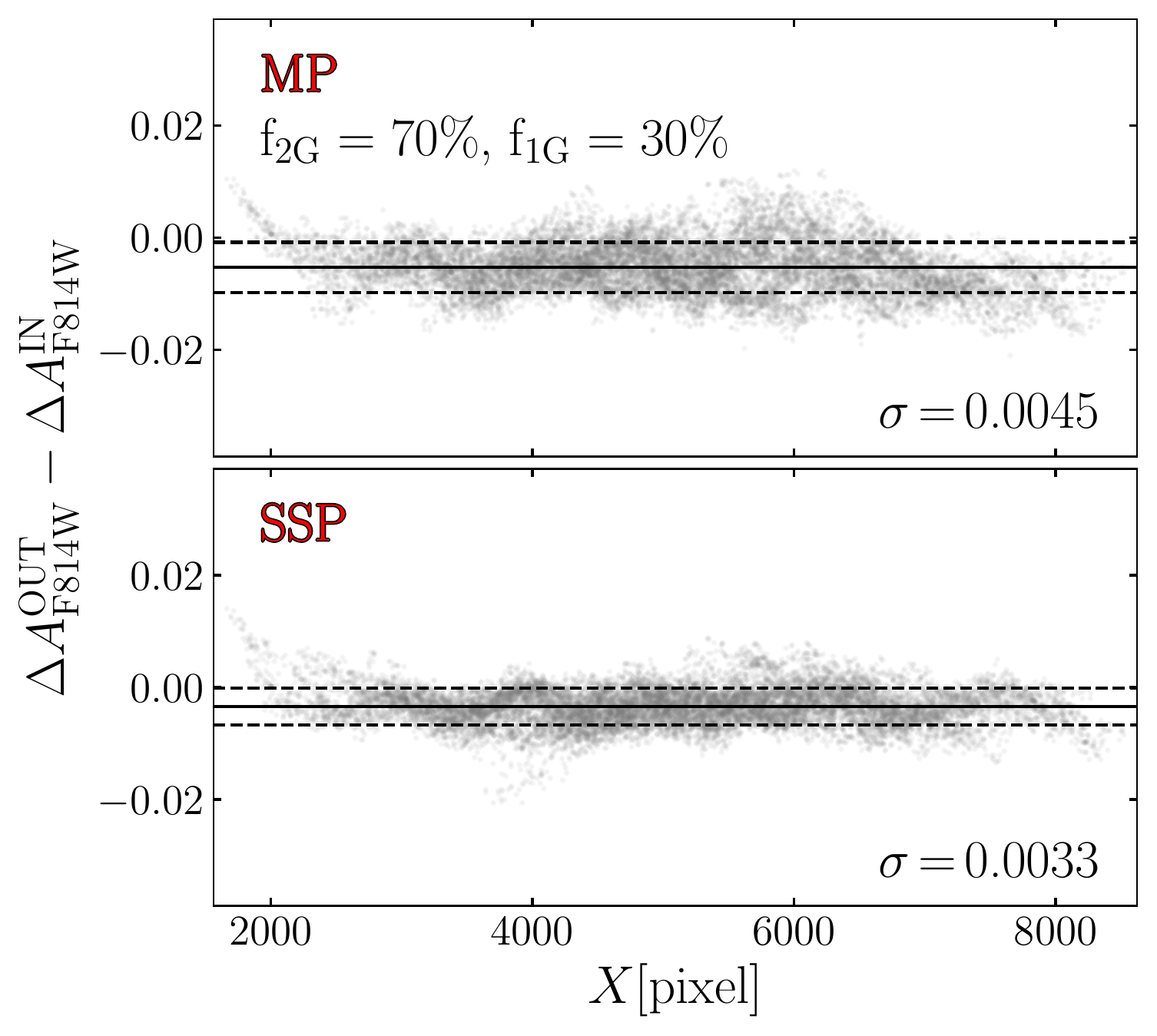}
    \caption{Difference between the measured differential-reddening values ($\Delta A_{\rm F814W}^{\rm OUT}$) and the values of differential reddening used in the simulation ($\Delta A_{\rm F814W}^{\rm IN}$) as a function of the X coordinates. The black continuous line marks the average difference, whereas the dashed lines are indicative of the dispersion of the residuals. We show the results in the case of a simulation of a GC with two stellar populations (top) and a single stellar population (bottom).} 
    \label{fig:res}
\end{figure}

Results are shown in Fig.\,\ref{fig:res} where we plotted this quantity as a function of the X coordinate for the two investigated cases, namely CMDs hosting a simple stellar population (SSP, bottom) and two stellar populations (MP, top). In both cases, we obtained a similar value for the average difference between $\Delta A_{\rm F814W}^{\rm OUT}$ and $\Delta A_{\rm F814W}^{\rm IN}$, which is close to zero. However, we note that in the MP simulations we obtain a dispersion of the $\Delta A_{\rm F814W}^{\rm OUT}-\Delta A_{\rm F814W}^{\rm IN}$ quantity ($\sigma$) that is $\sim1.25$ times higher than that derived from the simple-stellar population CMDs. We conclude that the multiple populations affect the uncertainties on differential-reddening determinations but do not introduce significant systematic errors.

\bsp	
\label{lastpage}
\end{document}